\documentclass[10pt]{article}

\usepackage{url,graphicx,threeparttable}

\usepackage[margin=2cm]{geometry}

\usepackage[final]{changes}

\usepackage{natbib}

\title{Empowering open science with reflexive and spatialised indicators}

\author{
J. Raimbault$^{1,2,\ast}$, P-O. Chasset$^{2,3}$, C. Cottineau$^{4}$, H. Commenges$^{2}$, D. Pumain$^{2}$,\\
 C. Kosmopoulos$^{2}$  A. Banos$^{5}$ \medskip\\
$^{1}$UPS CNRS 3611 ISC-PIF, Paris, France\\
$^{2}$UMR CNRS 8504 G{\'e}ographie-cit{\'e}s, Paris, France\\
$^{3}$LISER, Luxembourg\\
$^{4}$UMR CNRS 8097 Centre Maurice Halbwachs, Paris, France\\
$^{5}$UMR IDEES 6266, Caen, Le Havre, Rouen, France\medskip\\
$^{\ast}$ Corresponding author: \texttt{juste.raimbault@polytechnique.edu}
}

\date{}

\begin{document}

\maketitle

\begin{abstract}
Bibliometrics have become commonplace and widely used by authors and journals to monitor, to evaluate and to identify their readership in an ever-increasingly publishing scientific world. With this contribution, we aim \deleted{to move from the near-real time counts }to investigate the semantic proximities and evolution of the papers published in the online journal Cybergeo since its creation in 1996. We \added{propose a dedicated interactive application that} compare\added{s} three strategies for building semantic networks, using keywords (self-declared themes), citations (areas of research using the papers published in Cybergeo) and full-texts (themes derived from the words used in writing). We interpret these networks and semantic proximities with respect to their temporal evolution as well as to their spatial expressions, by considering the countries studied in the papers under inquiry (Cybergeo being a journal of geography, most articles refer to a well-defined spatial envelope). Finally, we compare the three methods and conclude that their complementarity can help go beyond simple statistics to better understand the epistemological evolution of a scientific community and the readership target of the journal.

\end{abstract}

\section{Introduction}

Faced with the increasing number of articles, journals and channels of publication used by researchers in a digital world with increasing open access, journals need tools to identify their readership and authors need this information to better reach their target audience, using the relevant keywords, vocabulary and citations. This paper suggests a set of complementary digital tools to tackle these needs \added{by journal themselves, empowering them through a reflexive analysis of their content and fostering open science through more transparency}. To perform their functions and provide useful insights, the tools need to meet three requirements: 1) to go beyond the usual citation metrics and to give semantic and network analytics directly from the scientific contents of the papers; 2) to situate sets of papers according to the semantic fields of their topic and their geography; 3) to identify significant variations in research topics that may be linked to the geographical origin of their authors or to the country they analyse. \replaced{This last point is especially interesting for scientific journals of geography.}{This last point is especially interesting for our first case of study, which is a scientific journal of geography.}

Since the seminal work of Thomas Kuhn in the early 1960s, the development of science studies has been based on three disciplinary pillars: history of science, philosophy of science, sociology of science. In the 1980s, political sciences contributed by focusing on the links between knowledge production and knowledge utilisation. This ``political turn'' began with the creation of the journal \textit{Knowledge} in 1979. Since the late 1990s, science studies have been affected by a ``spatial turn'' and eventually, a geography of science emerged \citep{livingston_spaces_1995, livingston_science_2003, withers_place_2009}. More recently, the conjunction of complexity-based approaches, networks science and big data have introduced a ``quant turn'', with systematic analyses of citation networks and automated mining of large textual corpora (see e.g. recent synthesis such as \cite{hicks2015bibliometrics} on research metrics or \cite{borner2015mapping} on maps of science).\deleted{ The present paper follows the last two trends, as we propose a spatialised bibliometrics approach.} \added{The emergence of a new highly interdisciplinary field is coined as a ``science of science'' by \cite{fortunato2018science}. The study of science by itself is indeed a crucial aspect for the production of scientific knowledge, also known in social sciences as reflexivity. This paper contributes to this effort by showing a proof-of-concept of reflexive corpus analysis methods and tools for electronic journals which are prone to such developments \citep{pumain1996cyberjournals}. We particularly insist on the two aspects of the heterogeneity of measures and of the spatial dimension, which we now contextualize.}
 
\added{First, measures and approaches in bibliometrics are highly multi-dimensional.} \cite{cronin2014beyond} attempt to provide an overview of the complex nature of the measure of scientific publications and the intrinsic multidimensional nature of knowledge production. \replaced{They}{Their article} provide\replaced{s} both recent technical contributions and a critical approach\replaced{, insisting}{. It insists} on the ``Janus-faced nature of metrics''.\replaced{This confirms}{ confirming} that reducing knowledge production to a few dimensions is not only wrong but dangerous for science. Studies in bibliometrics which have the complementarity of different approaches as their main focus are rather rare. \added{\cite{wen2017mapping} construct maps of hydrological science by combining different types of networks such as citation networks and keyword networks and show the complementarity of these entries. Part of the difficulty arises from the disciplinary context.} \cite{2016arXiv160106075O} show that taking into account citation and discipline data into a multilayer network is useful to understand patterns of interdisciplinarity.

\added{Secondly,} the geographical dimension of science has \added{also} been studied by \replaced{several}{numerous targeted} studies.\deleted{such as (Maisonobe et al., 2016)}
\deleted{ who investigate collaboration patterns aggregated at the level of cities, or (Maisonobe et al., 2017)}
\deleted{ who show the geographical deconcentration of scientific success in terms of citation received.} \added{\cite{frenken2009spatial} propose a specific research program for the emerging field of \emph{spatial scientometrics}, including specific questions such as the spatial distribution of citations and activities, but also specific methodological issues linked to noise in spatial data or more classical geographical issues such as the modifiable aerial unit problem. The work presented in the following will thus focus on corpuses with a spatial dimension.}
  
The 20-year anniversary of the first digital-only journal in \replaced{geography}{social sciences}~\citep{pumain2001cybergeo,kosmopoulos2002cybergeo}, namely Cybergeo (\url{http://journals.openedition.org/cybergeo/}), was the occasion to analyse a consistent corpus of over 700 articles published in 7 languages, with respect to its semantics as well as to the geography of its authorship and readership. We performed a quantitative epistemology analysis of the scientific papers published since 1996 to measure their similarities according three types of textual indicators: their keywords (the way authors advertise their research), their citation network (the way the paper is used by other fields and disciplines), and their full-text (the vocabulary used to write the paper and present the research).

These analyses are complementary and show the evolution of a journal towards emergent themes of research. It also highlights the need for Cybergeo to keep extending its authorship base beyond the French-speaking community, in order to match its ambition to be a European Journal of Geography. Our contribution \replaced{mainly consists in a }{consists in these specific epistemological conclusions, but also in a broader} methodological and technical product developed to interactively handle a large-scale heterogeneous scientific corpus\added{, with a particular attention to spatialized corpuses}. We show how the coupling of complementary views can create a second-order type of \replaced{knowledge on the scientific context of the corpus studied}{epistomological knowledge}: the spatial embedding of the three classification methods unveils unexpected patterns. Furthermore, the dedicated online tool that we designed is available as an open source software which can be used by journals for a collective scientific reflexivity, but also by institutions and individual scientists for a bottom-up empowerment of Open Science.

The \replaced{remaining}{rest} of the paper is organised as follows.\deleted{: we first review similar initiatives tackling heterogeneous or multidimensional approaches to bibliometrics, and describe the case study.} \replaced{We first}{We then} describe the different methods used to analyse semantic networks, and how these are coupled through interactive spatial data exploration. \replaced{We then}{Finally, we} describe results at the first-order (each method) and at the second-order (achieved through coupling) before discussing broader implications for quantitative epistemology and reflexivity in Open Science.

\section{\replaced{A multi-method spatialized corpus analysis tool}{Methods}}

One main aspect of our contribution is the combination of different methodologies, each having its potentialities and pitfalls, but also specific questions and objects of study. We present in this section the different methods and how they are coupled together to produce second-order knowledge.

\subsection{Internal semantic network}

The first exploration method is based on the set of keywords declared by the authors themselves when publishing in Cybergeo. We consider articles and keywords as a bipartite network. This network can be decomposed in two simple networks: a network of articles (vertices) linked by common keywords (edges); a network of keywords (vertices) linked by common articles (edges) \citep{roth_social_2010}. We consider the second one as a semantic network. \deleted{Based on a measure of modularity,} We construct semantic communities (see supplementary material for methodological details) with the Louvain algorithm \citep{blondel_fast_2008}. This community detection method is chosen among others because it is based on modularity measures such as the modal weight defined above. \added{The Louvain method performs a modularity maximisation, such as other algorithms. In this case the semantic network is small and simple and any modularity-based algorithm would give a similar output.}

\subsection{External semantic network}

The second methodological development focuses on the combination of citation network exploration and semantic network analysis. The full method we apply here is described in details by~\citep{raimbault2017exploration}. Citation networks have been widely used in science studies, for example as a predictive tool for the success of a paper~\citep{2013arXiv1310.8220N}, or to unveil emerging research fronts~\citep{shibata2008detecting}. Indeed, the bibliography of a paper contains a certain scientific positioning, as well as a line of inheritance to which it aims to contribute and which fields it is based on. Reverse citations (i.e. contributions citing a given paper, up to a given level) on the other hand show how the knowledge presented in a paper was understood, interpreted and used, and in particular by which field (on this point the interesting example of \deleted{Jane}\citep{jacobs1961death}, heavily cited today by quantitative studies of the city by physicists, shows how unexpected the audience can be over time).

We define the citation neighborhood of our corpus as all the articles citing articles published in \textit{Cybergeo}, all the articles citing the ones cited by \textit{Cybergeo}, and all the articles citing these ones. We therefore have a network of depth 2, with a control group to compare to Cybergeo articles. The citation data is collected using automatic data collection \citep{raimbault2017exploration}. Once citation neighborhoods have been constructed, keywords are automatically extracted from the abstracts of corresponding publications using natural language processing techniques, and a semantic network is constructed (see supplementary material for details). This network and its communities \replaced{enable}{allow us} to associate a list of keywords and corresponding disciplines to each paper. These are complementary to the declared keywords and the full-text themes presented in the next subsection, as they reveal how authors position their article in the semantic landscape associated to the citation neighborhood, or what their ``cultural background'' is.

\subsection{Topics allocation using full-text documents}

The third and last exploration method details the allocation of topics in full text documents, and is thus complementary to the previous ones that used declared keywords and relevant keywords within abstracts of the citation neighborhood. Topics allocation is used widely for many purposes. \cite{CITRON2018181} have identified semantically-related scientific communities. \cite{KARAMI20181} have characterized diabetes, diet, exercise, and obesity in comments on Twitter. \cite{niebles_unsupervised_2008} have found human action categories in video sequences translated in spatiotemporal words. All these authors have mobilized a topic classification model such as the Latent Dirichlet Allocation model (LDA) and its derivatives. We apply here the LDA method to extract topics from full-text documents (see supplementary material for a thorough description of the method).

\subsection{Geographical aggregation of semantic profiles}

Given a semantic characterisation of articles (using keywords, citations or full-texts), it is then possible to determine two semantic profiles of countries: one using countries as authoring 'origins' and one using countries as subject 'destination'. This semantic profile of a country X is made of the mean share of themes Y present in articles authoring from or studying country X. At one extreme, if only one article $A_1$ came from a country $X_1$, the semantic profile of $X_1$ would be exactly that of $A_1$. At the other extreme, if all articles came from $X_2$, the semantic profile of $X_2$ would be the overall distribution of themes across the corpus.

All in all, given the three semantic characterisations of articles (using keywords, citations and full-texts) and the two geographical allocation of articles (authoring or studied), each country has a maximum of six distinct semantic profiles. We use these semantic profiles to cluster countries. The clustering method applied is an ascending hierarchical clustering algorithm using the Ward criterion of distance maximisation. When analysing authoring clusters, we consider groups of countries from which a certain geography is made and written. This option is interesting in a reflexive aspect but practically more hazardous because of the high concentration of emissions (and the consequently low number of emitting countries) and because of the uncertainty of national provenance as captured by the institutional affiliation of authors at the time of publication. Therefore, in the \replaced{application}{results} section, we base our clustering on studied countries only. When analysing clusters of studied countries, we consider how certain groups of territories are studied, what words authors use to talk about them and in which research areas the papers about them are used.

\subsection{Open data + interactivity = reproducibility \& transparency}

Last but not least, our methodological contribution is also closely linked to issues of reflexivity, transparency and reproducibility in the process of knowledge production. It is now a well sustained idea that all these aspects are closely linked and that their strong coupling participate in a virtuous circle enhancing and accelerating knowledge production, as seen in the various approaches of Open Science~\citep{fecher2014open}. For example, open peer review is progressively emerging as an alternative way to the rigid and slow classical canons of scientific communication \citep{10.12688/f1000research.11369.1}. In the domain of computational science, tools are numerous to ensure reproducibility and transparency but require a strict discipline of use and are not easily accessible~\citep{wilson2017good}. Open Science suggests transparency of the knowledge production process itself, but also of the knowledge communication patterns: on this point we claim that the interactive exploration of quantitative epistemological patterns is necessary. We therefore built an interactive application to \replaced{enable}{allow} the exploration of heterogeneous scientific corpora.

The web application is available online at \url{http://shiny.parisgeo.cnrs.fr/CybergeoNetworks/}. Source code and data, both for analyses and the web application, are available on the open \texttt{git} repository of the project at \url{https://github.com/AnonymousAuthor3/cybergeo20}.


\section{\replaced{Application to the Cybergeo corpus}{Results}}

\subsection{\replaced{Data}{Cybergeo as a case study}}

\added{The data used to test our method and tools consist in the corpus of 20 years of article publication from a free online geography journal.} Cybergeo was founded in 1996 as a digital-only European journal of geography. Between April 1996 and May 2016, 737 scientific articles have been published by 1351 authors from 51 countries. These articles have generated 2710 citations altogether over the last twenty years, which correspond to half the number of all the other articles cited by Cybergeo articles (5545).

In order to produce 
 analyses at the country level, articles have been geo-tagged in two ways. Firstly, the country of affiliation \replaced{as it was declared by the author(s) at the moment of publishing}{of the author(s)} has been coded following the 2-letter identifiers of the International Organization for Standardization. \added{This information is available as authors have to choose a single or major affiliation and fill a form on the publishing platform.} Secondly, the articles were read one by one to extract the major geographical subjects. Articles were tagged with a country if this country or a sub-region of it constituted the focus of the study. In the case of European countries, different sets of countries were associated with the publication, depending on the perimeter of the subject (for instance: EU15, EU25, Schengen area, EuroMed, etc.).

\begin{figure}
	\includegraphics[width=\linewidth]{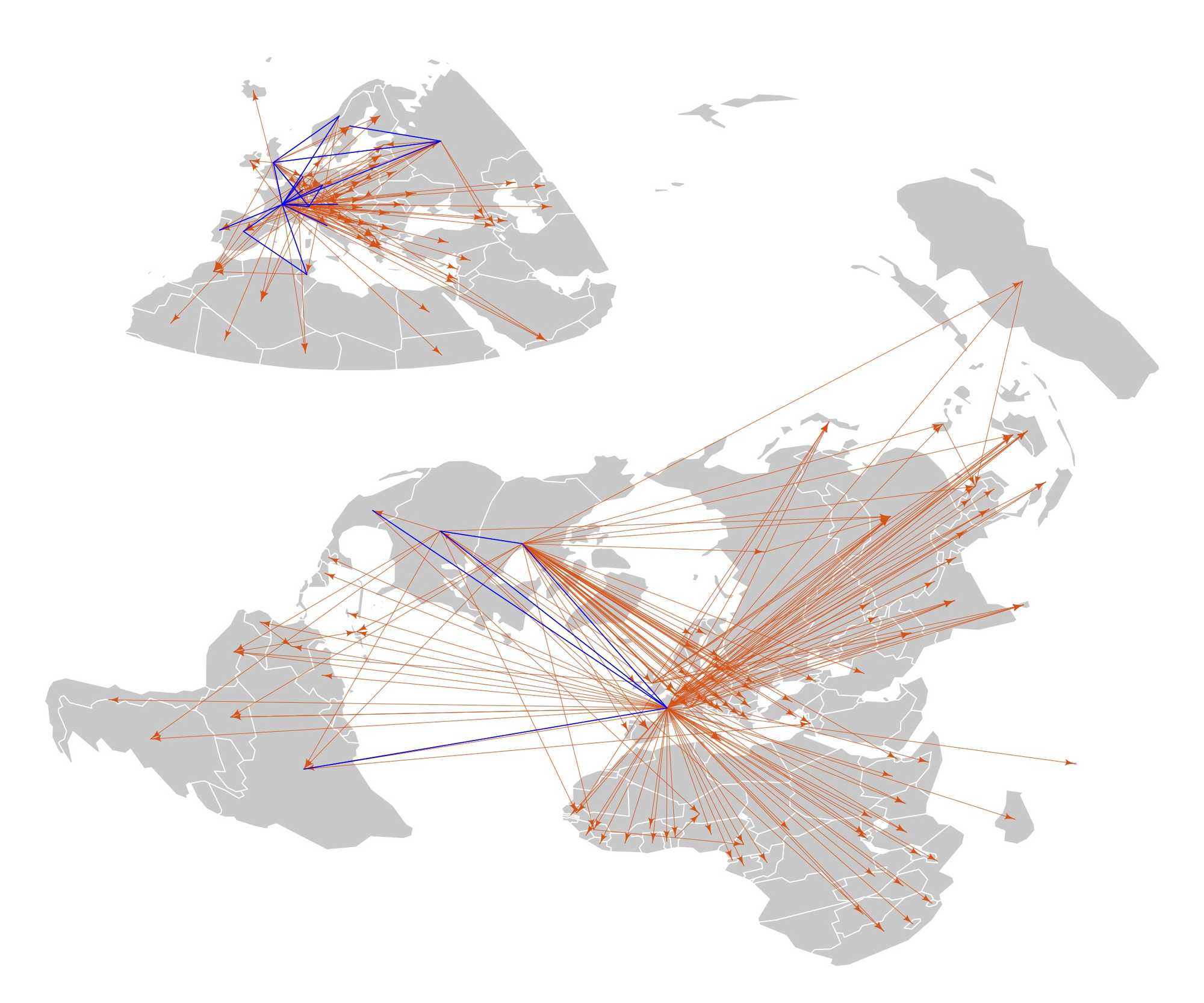}
\caption{Geographical origins and destinations of papers | 1996-2015. Reciprocal links are represented in blue.} 
\label{fig:who} 
\end{figure}

Summary statistics of authorship by country are given in supplementary material. By linking institutions of authors to their geographical subject (fig. \ref{fig:who}), we find different patterns:
\begin{itemize}
\item European and North American countries tend to study each other in a symmetric way through Cybergeo articles;
\item Latin American countries are mainly studied by authors affiliated in Europe and North-America;
\item African and Asian countries are studied mainly by Europeans and marginally by Americans and themselves;
\item Russia and Australia are studied by Western authors and study their own hinterland.
\end{itemize}
\added{Finally, we find privileged links between France and (formerly) French-speaking countries (including Belgium, Canada, Vietnam, Madagascar, Senegal etc.),  whereby a common language and at times a shared history through colonialism have produced favoured national subjects of study to be published in the European (yet predominantly French-speaking) Cybergeo journal.}

\subsection{Internal semantic network (keywords)}
\subsubsection{Communities and semantic fields}

\begin{figure}
	\includegraphics[width=\linewidth]{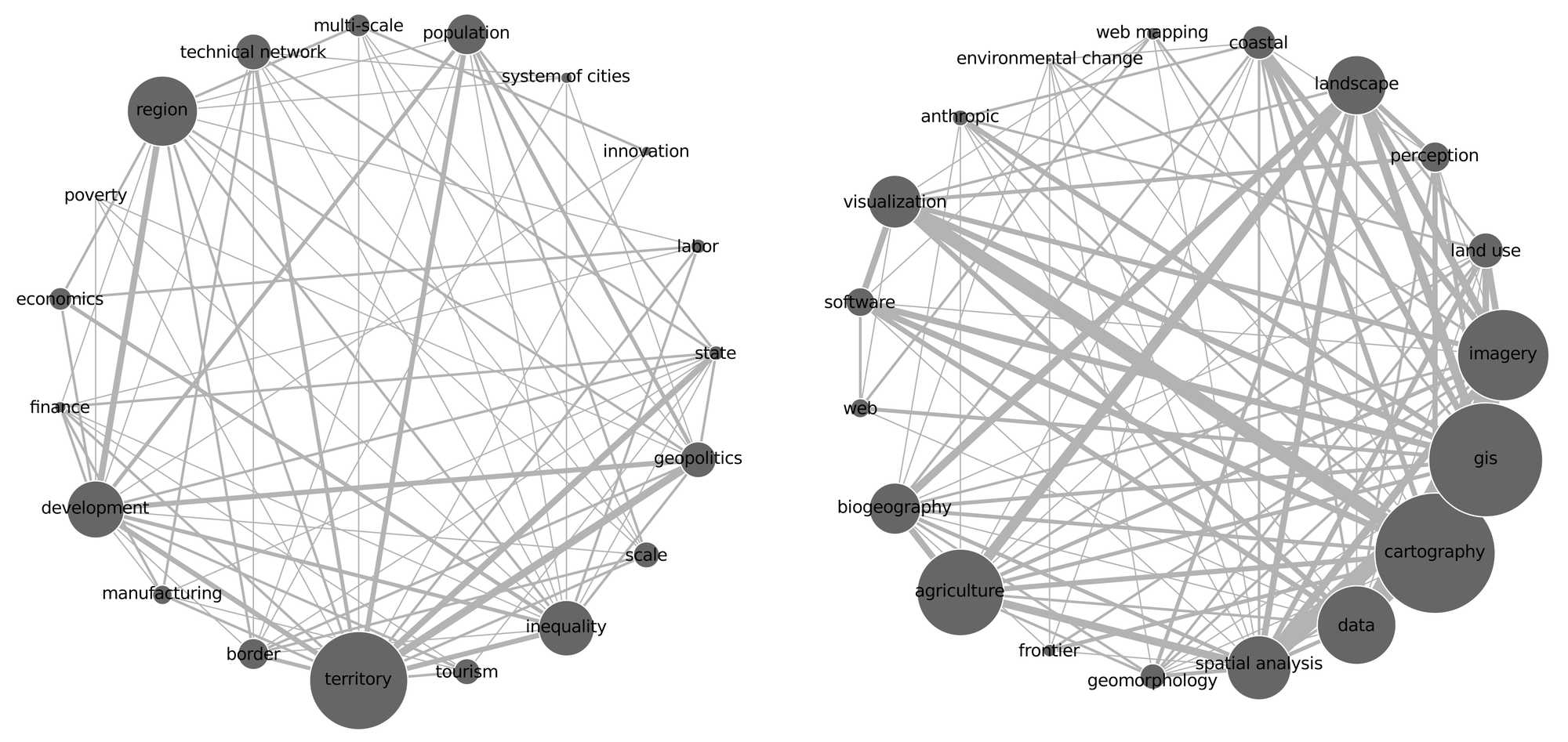}
\caption{Community structure of the internal semantic network (A-Territory; B-Imagery \& GIS).}
\label{fig:commintern}
\end{figure}

The community detection algorithm reaches \deleted{its modularity optimum }a modularity optimum with 10 clusters\added{, which we summarize by expert knowledge (as we will do for each clustering result in the following) as}: mobility and transportation; imagery and GIS\footnote{short for Geographical Information Systems}; climate and environment; history and epistemology; sustainability, risk, planning; Economic geography; Territory and population; urban dynamics; statistics and modelling; emotional geography. Some clusters concentrate a large number of keywords and articles, such as "imagery and GIS" or "statistics and modelling". This result was expected because of the original aim and scope of the journal (quantitative geography). Beside the main clusters and a set of medium-sized clusters, two small and totally unexpected clusters \added{according to journal editors} emerged: "emotional geography" and "climate and environment". The CybergeoNetworks application proposes a set of visualisation parameters to draw the communities (see Figure~\ref{fig:commintern}) such as setting the size of vertices and edges according to different variables (degree, number of articles, modal weight).

\begin{figure}
	\includegraphics[width=\linewidth]{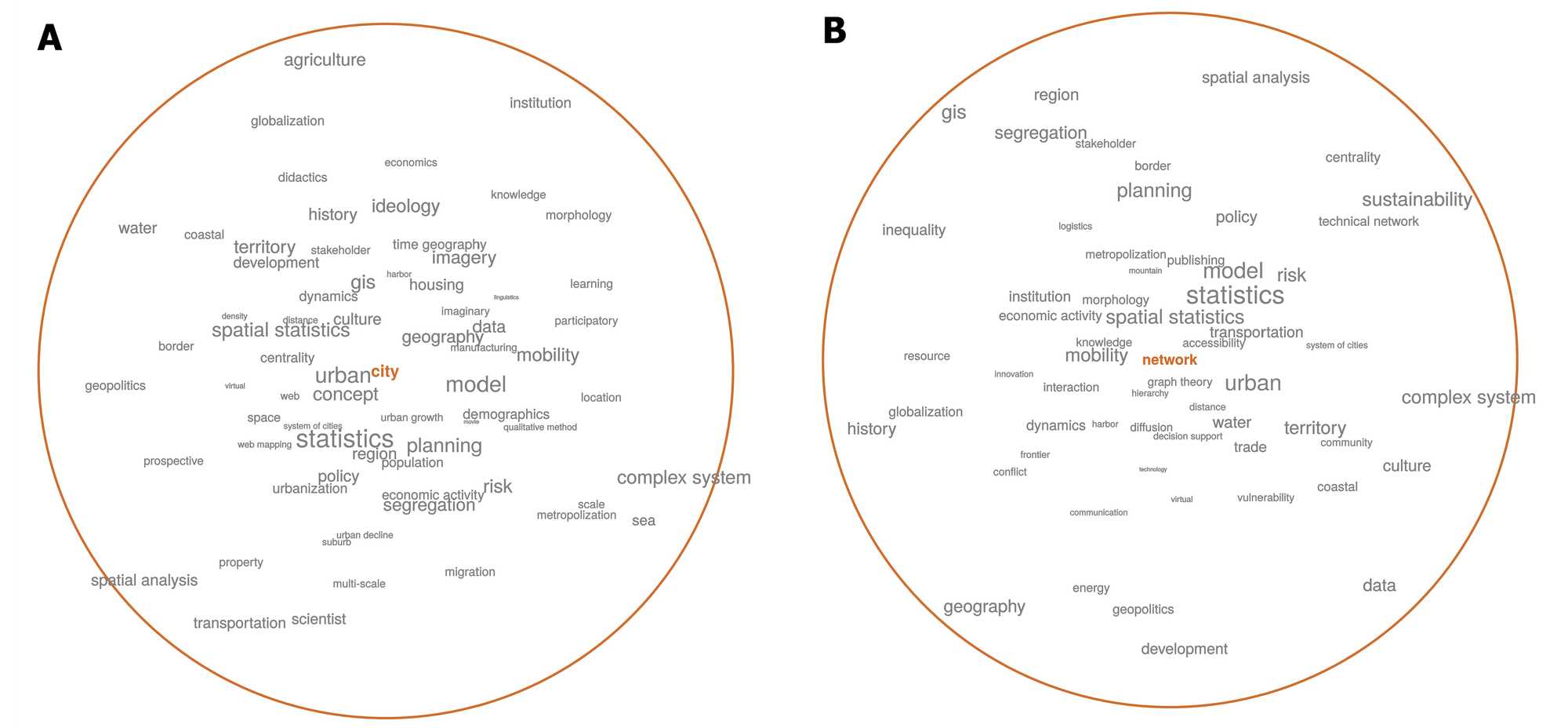}
\caption{Semantic fields (A: "City"; B: "Network").}
\label{fig:fieldintern}
\end{figure}

As explained above, modal weight metrics can be used to draw semantic fields. The CybergeoNetworks application presents the full list of keywords. The user chooses one keyword from that list, the word is placed at the centre of the plot and all its neighbours are arranged at a distance inversely proportional to the preferential attachment (modal weight). We illustrate this feature in figure~\ref{fig:fieldintern} : to ease reading, a circle is drawn at a distance of 1. Some proximities are expected ("urban" is closely linked to "city"), some are expected knowing the original scope of the journal in the field of theoretical and quantitative geography ("model" or "spatial statistics" are linked to "city"). Some proximities are totally unexpected: for "city", the preferential attachment of keywords like "movie", "web", "virtual".

\subsubsection{Spatial communities}

\begin{figure}
	\includegraphics[width=\linewidth]{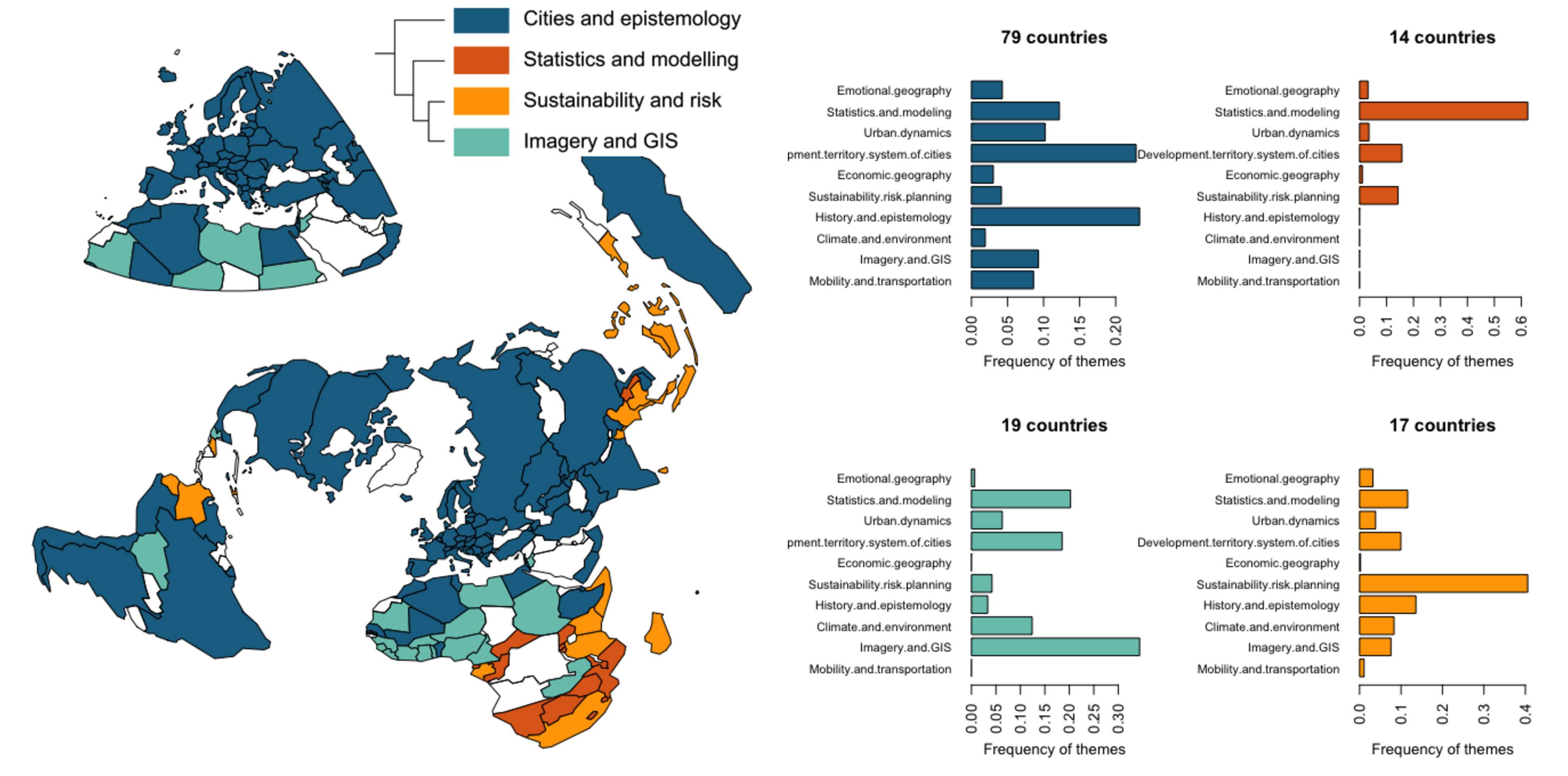}
\caption{\textit{(Left)} Geographical communities of declared interest. \textit{(Right)} Corresponding semantic profile of groups.} 
\label{fig:cluster_hadri} 
\end{figure}

Using the keywords distributions to draw the semantic profile of the 129 countries studied in a Cybergeo article, we obtain a clustering in 4 groups representing 16.3\% of the initial inertia\footnote{a relatively low level, although a clear-cut in dendrogram allowing a manageable number of groups to observe the main differences between countries}. Its geographical distribution is shown in figure \ref{fig:cluster_hadri} with the average profile of each group.

Countries are differentiated firstly by whether or not the articles studying them also declare keywords related to transport and mobility, history and epistemology, urban systems and/or emotional geography. Indeed, the first group of 79 countries (in blue, figure \ref{fig:cluster_hadri}) is defined by these themes. The corresponding countries are the most developed and richest territories of the world, including emergent countries such as the BRIC\added{S countries (Brazil, Russia, India, China and South Africa)}. The keywords used to advertise the articles follow the latest trends of geography, with mentions of emotions and mobility for instance. 

The countries of the other groups over-represent the keywords related to:
\begin{itemize}
\item methods (in orange) such as statistics and modelling. The countries associated with these keywords are all located in central and southern Africa, with the exception of Lao\added{s}. These countries are studied by a small number of articles focusing on methodological approaches. For example, the only article studying Rwanda \citep{querriau2004localisation} relates to an optimal location problem whereas \cite{vallee2009disparites} uses 'multilevel modelling' as a keyword for the only article about Lao.
\item sustainability and risks (in yellow). This is the case of articles about Indonesia for example, which all relate to hazards and vulnerability: to tsunamis \citep{ozer2005tsunami}, to volcanoes \citep{belizal2011quand} and to water scarcity \citep{putra2009einfluss}.
\item Finally, 19 countries are associated with keywords related to imagery and GIS (teal colour). They are located primarily in Saharan Africa. In many cases, this happens because the articles present a methodology which uses aerial and satellite images to substitute missing socioeconomic data  \citep{ackermann2003analysis, devaux2007extraction}.
\end{itemize}

Thus, drawing communities of declared interest, we find an interesting dichotomy between rich countries on the one hand, which are studied extensively in the literature and for which authors use trendy keywords to singularise themselves from past and concurrent work; and developing countries on the other hand, which are associated with more technical keywords reflecting a narrower spectrum of domains and specific data challenges.

\subsection{External semantic network (citations)}

The application \replaced{enables to explore}{allows exploring} the citation neighborhood of chosen articles, in terms of semantic contents (the visualisation of full networks are technically not feasible as the full corpus contains around 200,000 articles). Wordclouds on the CybergeoNetworks application give the content of the article and the content of the articles in the neighborhood, with each word being associated to the semantic communities. The user can therefore situate a work within a semantic context, and we expect that unanticipated connexions can be made with this tools, as authors may not be aware of similar works in other disciplines.

\subsubsection{Communities structures}

As explained before, the raw semantic network is optimized for modularity and size, making a compromise between these two opposite objectives, when edge and node filtering parameter vary. This provides 12 communities, which can correspond to existing disciplines, to methodological issues, or to very precise thematic subjects. The communities are, in order of importance in terms of proportion of total keywords: Political Science/Communication; Biogeography; Social and Economic Geography; Climate; Physical Geography; Commerce; Spatial Analysis; Microbiology; Neuroscience; GIS; Agriculture; Health. This method has the property of grouping together keywords based on co-occurrence, thus revealing the actual structure of abstracts' contents: it is both an advantage when revealing links as for the large field of Social and Economic Geography, but it can also blur information by grouping more detailed communities. Very precise and small communities such as Health Geography appear as they are strongly isolated from the rest of the communities. This structure is particular, and shows a dimension of knowledge that classical citation analysis would not reveal.

\subsubsection{Spatial communities}
\begin{figure}
	\includegraphics[width=\linewidth]{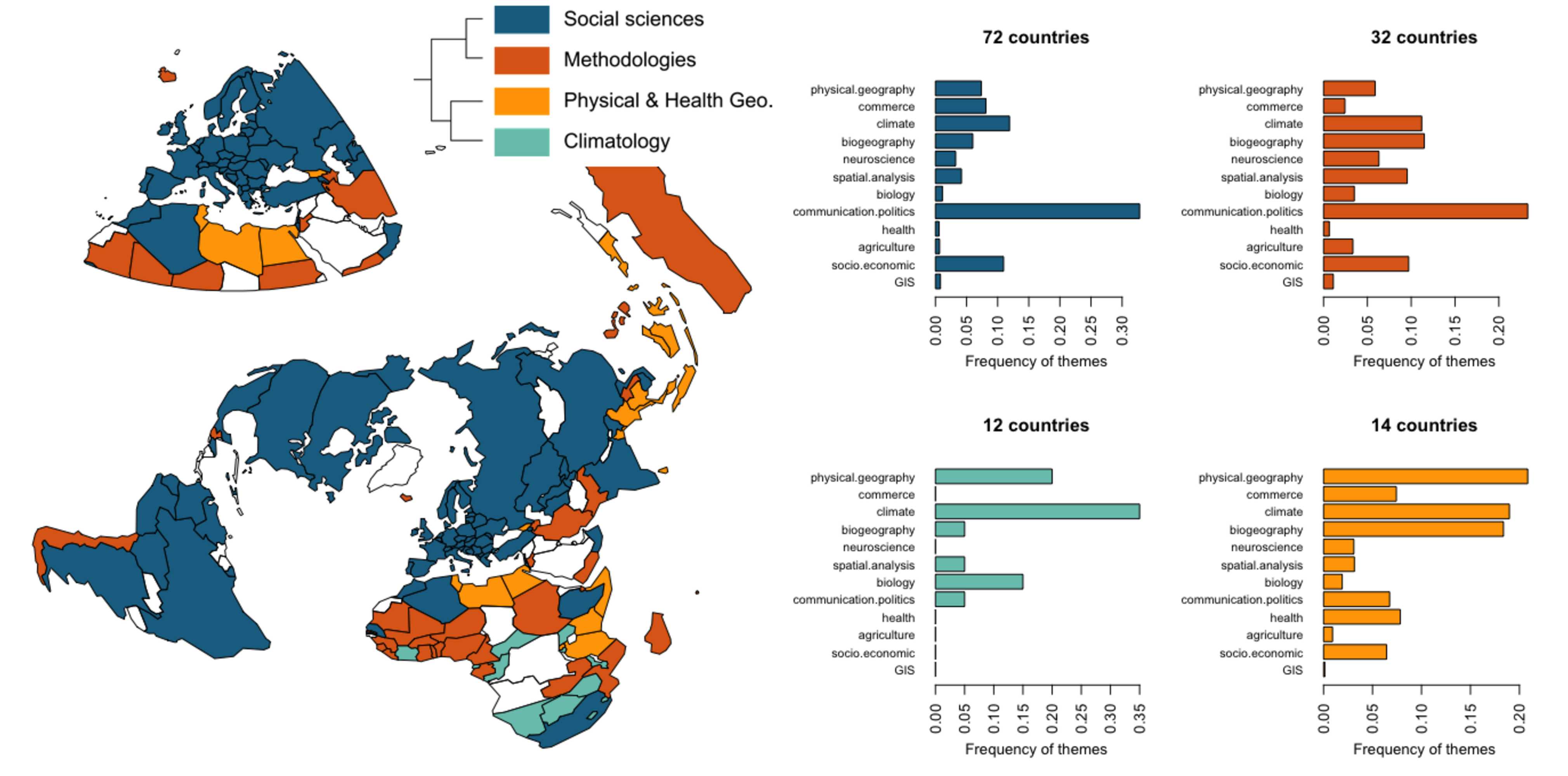}
\caption{\textit{(Left)} Geographical communities of bibliographical use. \textit{(Right)} Corresponding semantic profile of groups.} 
\label{fig:cluster_juste}
\end{figure}

Using the citation network communities to draw the semantic profile of the 130 countries studied in a Cybergeo article, we obtain a clustering in 4 groups representing 16.4\% of the initial inertia\footnote{\added{The results for other numbers of classes can be produced using the dedicated interactive online tool CybergeoNetworks.}}. Its geographical distribution is shown in figure \ref{fig:cluster_juste} with the average profile of each group.

The largest group of countries largely overlaps with the largest cluster of keywords communities (cf. previous section). Indeed, rich and emergent countries (BRICS included) are studied in articles used in similar ways in the citation network. There are further divides among this group. A first subgroup (in blue) of countries is studied by Cybergeo articles cited preferentially in the fields of commerce, socio-economic and political analysis. These correspond to articles mostly in Economics and Social Sciences. The nearest subgroup of countries (in orange) comprises 32 'Southern' countries such as Australia, Pakistan, Chile, Madagascar, Iran, Lao, the Philippines or Iceland. It corresponds to countries treated by articles cited preferentially in methodological fields (spatial analysis and GIS). Indeed, the only article about Iran presents a collaborative decision support system \citep{jelokhani2012web} while the only article about Australia reviews online cartographic products \citep{escobar2000distribution}. This kind of articles then tends to stay in the citation clique of geomatics. The third refers to 14 countries in South-East Asia (Indonesia, Thailand, Myanmar), Eastern Africa (Somalia, Kenya, Tanzania) and North Africa (Libya, Tunisia, Egypt). The articles studying them are cited preferentially in the fields of physical geography and health studies. The vision of these countries through the articles citing works published in Cybergeo is thus dominantly one of morphological wonders and health vulnerability. Finally, a group of 12 sub-Saharan countries (C{\^o}te d'Ivoire, Zimbabwe, the Republic of Congo) are associated with papers cited in the climatology citation community. 

Thus, drawing communities of bibliographical use, we find an interesting  dichotomy between rich countries on the one hand, which are associated with papers cited in broad communities, including topical and methodological fields; and poor and developing countries on the other hand, which are associated with papers cited mainly in relation to their natural geography, health and climatic risks in the literature. This could suggest a need for the journal to call for more articles about such countries' populations and economies.

\subsection{Topics allocation (full-texts)}

\subsubsection{Evolution of the topics addressed in the corpus}

\begin{figure}
	\includegraphics[width=\linewidth]{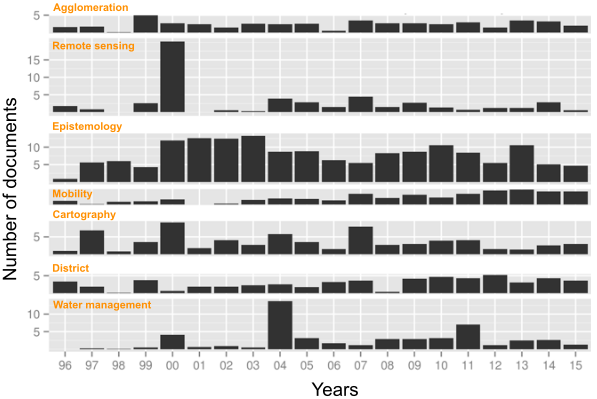}
\caption{Number of documents addressing a topic per year, between 1996 and 2015\added{. For visualization purposes, only selected topics commented in text are shown.}} 
\label{fig:topics-evolution} 
\end{figure}

The \added{Latent Dirichlet Allocation (LDA)} model is applied on a reduced corpus of French articles, this language being the leading one for Cybergeo. \added{We chose not to translate articles and keep a larger corpus with French articles, in order to avoid potential additional bias due to the translation process in our results.} After destructuring the texts and filtering nouns, articles and verbs, our corpus counts no less than four million words, which leads to a dictionary of 137,224 unique words. The optimal number of topics was chosen by estimating the LDA parameters for different numbers of topics and choosing a compromise between perplexity and entropy of the resulting classification (see details \added{of this optimization and the detailed description of topics} in supplementary material), what results in 20 topics. We give in supplementary material a part of the matrix describing, for each topic index, the first 20 translated words (except for the index 7 where words were already in English) in decreasing order of probability to belong to the topic. It is then interesting to observe how many documents addressed a given topic each year, i.e. the topics evolution in the Cybergeo corpus (fig. \ref{fig:topics-evolution}). We can distinguish several evolution profiles: decreasing, punctual, regular and increasing topics. Articles about cartography \deleted{(11) }tends to decrease. Articles about remote sensing \deleted{(3) }were mainly produced in 2000, just like articles about water management \deleted{(10) }in 2004 and 2011. Articles about agglomeration \deleted{(14) }are regularly produced. \added{Geographical epistemology is also often debated across articles.} Topics such as district \deleted{(1) }and mobility \deleted{(2) }tends to increase.

\subsubsection{Spatial full-text communities} 

\begin{figure}
	\includegraphics[width=\linewidth]{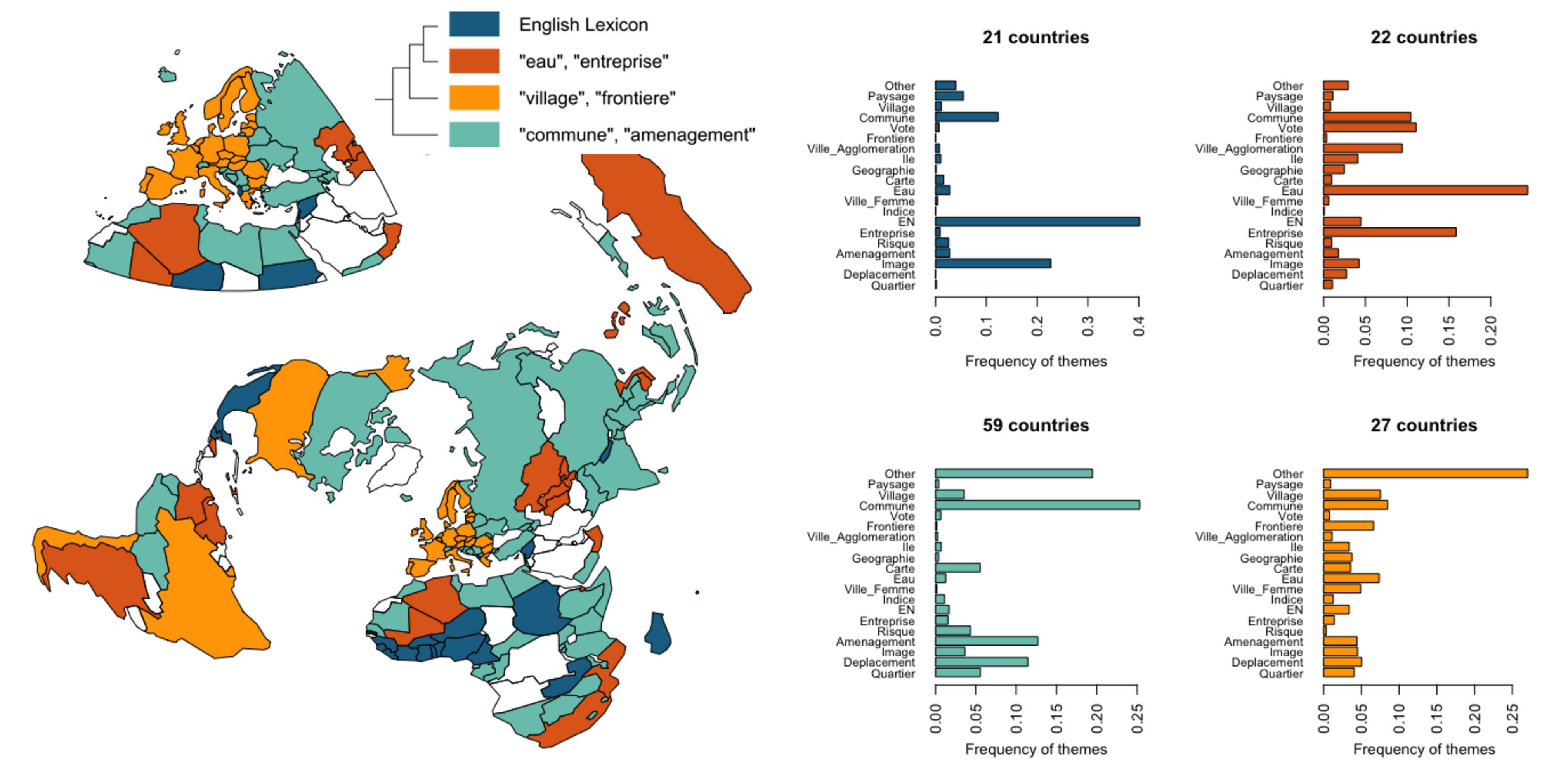}
\caption{\textit{(Left)} Geographical communities of writing practice. \textit{(Right)} Corresponding semantic profile of groups.} 
\label{fig:cluster_poc}
\end{figure}

Using the full-texts to draw the semantic profile of the 129 countries studied in a Cybergeo article, we obtain a clustering in 4 groups representing 13.4\% of the initial inertia. Its geographical distribution is shown in figure \ref{fig:cluster_poc} with the average profile of each group.

In this clustering analysis, we do not find the dichotomy of countries based on their wealth and economic development levels. The link between semantic and geographical proximity is also less obvious at the world level, \replaced{in spite of one region being}{although one region is} strikingly revealed: the institutional boundaries of Europe. The group of countries included in the EU27 plus the USA, Brazil and Chile (in yellow) appears strongly similar in terms of vocabulary used to talk about them. In particular, themes related to borders ("\textit{fronti{\`e}re}") and villages describe these countries well (for example: \citep{santamaria2009schema, lusso2009musees, le2011consommation}). 

A second group includes countries studied by papers written in English, as Cybergeo is a multilingual journal. A group of 59 countries, including Canada, Russia, Namibia, Malaysia and Ecuador, are studied in Cybergeo using preferentially words such as "commune", "d{\'e}placement" (mobility or displacement) and "am{\'e}nagement" (planning), suggesting an effort from French speaking authors to present and explain the geographical and urbanism context of other countries around the world. Finally, a group includes countries from all continents and corresponds to papers written preferentially with words such as "\textit{eau}" (water) and "\textit{entreprise}" (enterprise), that is a very heterogeneous set of papers which could easily be classified as "other".

The communities of vocabulary and writing practice thus appear less straightforward and less linked to geographical proximity. The main result lays in the fact that there is a specific set of words used to write about the European Union, a sort of EU27 Novlang made of words like "Eurovision", "subsidiarity" and "Spatial Development Perspectives".

\section{Discussion}

\subsection{Why three classifications? Evaluating the complementarity of approaches}

This section backs up the previous qualitative comparison of approaches through their spatialization by quantitative measures of their complementarity. \replaced{In spite of having}{Although we have} seen that the communities obtained from the three different methods are semantically and geographically distinct, we do not know precisely how they complement each other. The overlapping analysis is complicated by the fact that articles belong simultaneously to several clusters for each classification. 

Therefore, we compare the methods 2 by 2 by computing the share of articles classified simultaneously in each possible pair of clusters from the two methods. Methodological details are detailed in supplementary material, together with the diagram synthesizing overlap between communities of the different classifications. We obtain for instance a clear preferential positive and negative relations between some citations communities and keywords communities (figure \ref{fig:complementarity}). On the one hand, 35\% of the Cybergeo articles in the GIS citation cluster are characterized by keywords identified as "Imagery and GIS". On the other hand, there is no article in the "crime" citation cluster which have keywords of the "Climate and environment" community. These relationships make sense, because the way a paper is advertised by its keywords is one of the first elements indicating the potential reader that the paper is relevant or not. Interestingly, the "complex systems" citation community is characterized by a variety of keywords communities (27\% of the articles cited by this community are tagged in the "statistics and modelling" cluster, 17\% in "Imagery and GIS" cluster, 13\% in "history and epistemology", 11\% in "urban dynamics"). This suggests that the field of complex systems, being unified by methods rather than objects of inquiry, are more open to diverse topics than other citation communities. It could also mean that within Cybergeo, authors of articles relevant to the complex systems community advertise their paper with keywords from the discipline of geography rather than methods only, in order to attract topical readers as well. Looking at the relations between keywords communities and themes communities, we find that some topics require specific words to write about them. For example, "Imagery and GIS"-tagged articles use more words from the "EN" theme category, which corresponds to English words (rather than French). Urban studies are distinguished between its quantitative side (advertised by keywords around "urban dynamics" and using words such as "agglomeration") and its qualitative side (advertised by keywords around "sustainability, risk and planning" and using words such as "\textit{femme}":  woman). Interestingly, the words like "risk" (\textit{risque}) are used themselves more in articles tagged around "Climate and environment" than around "sustainability, risk and planning".

\begin{figure}
	\includegraphics[width=\linewidth]{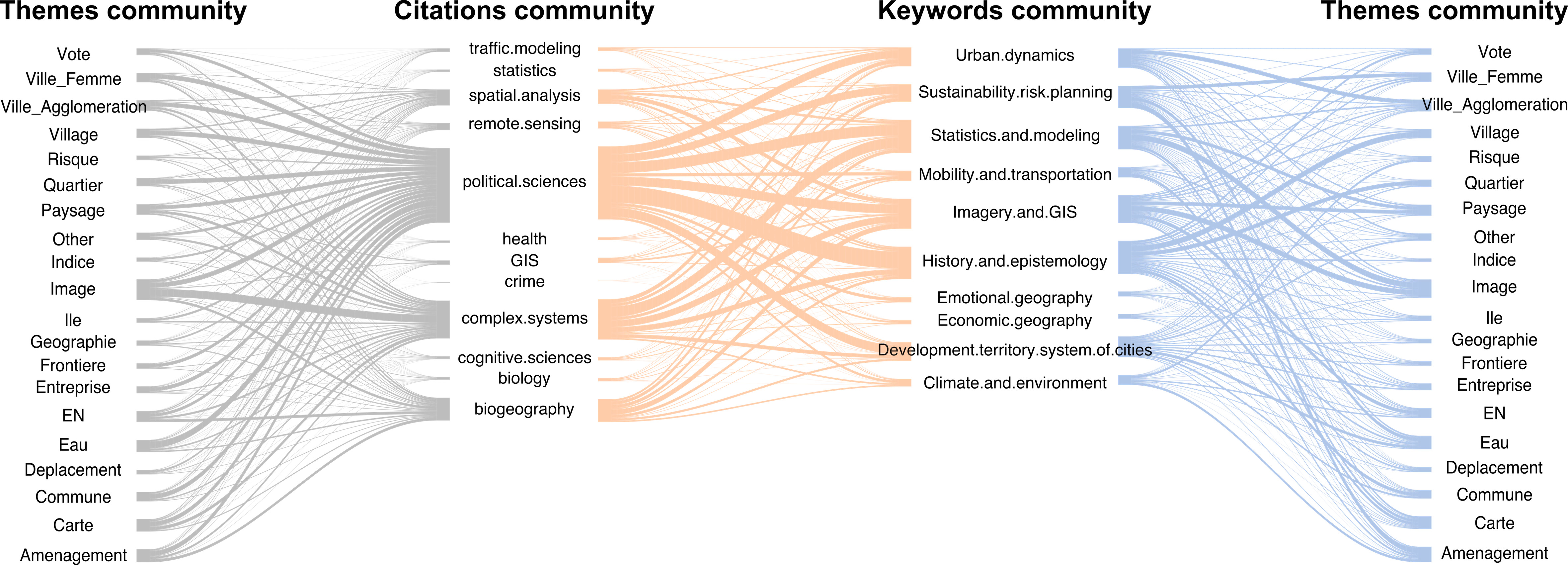}
\caption{Overlapping between the three methods of semantic classification} 
\label{fig:complementarity}
\end{figure}

Finally, the flows between themes communities and citations communities appear roughly proportional to the size of clusters at origin and destination, suggesting that citations are rather independent of the vocabulary used in the articles. This is reflected in the quantitative analysis of correlations done in supplementary material 
, this pair having the smallest mean absolute correlation. In short, the words that count in a citation strategy are much more the keywords than the actual content of the paper. These complementary analyses show thus the complementarity of classifications in the exploration of semantic diversity of publication in a 20 year old journal.


\deleted{Our approach can be understood as an ``applied perspectivism'', which we believe is a way to enhance second-order knowledge creation and to ensure reflexivity. Perspectivism is an epistemological position defended by (Giere, 2010)}
\deleted{, which aims to go beyond the constructivism-reductivism debates. Focusing on scientific agents as carriers of knowledge creation, any scientific enterprise is a certain \emph{perspective} on the world, taken by the agent for a given purpose and through a given medium that is considered as the \emph{model}. Perspectives are necessarily complementary as they result from different approaches to the same objects, even if the definition of objects and research questions will not necessarily be the same. Coupling perspectives should thus be a typical feature of interdisciplinarity. We position our work as a deliberate attempt to couple complementary perspectives on the same corpus. }
\deleted{Varenne (2017) recalls that one of the various function of models is to foster coupling between theories through coupling of models themselves, allowing the creation of novel knowledge within the virtuous spiral between disciplinarity and interdisciplinarity coined by (Banos, 2017).}
\deleted{Our work aims precisely at accelerating and improving such processes.}

\section{\replaced{Conclusion: Fostering Open Science and Reflexivity}{Fostering Open Science and Reflexivity}}

The open tools and software we provide participate to a larger effort of reflexivity tools in the context of Open Science. It is aimed at being complementary to existing platforms, like the Community Explorer for the community of Complex Systems developed by ISCPIF\footnote{available at \url{https://communityexplorer.org}} that provides an interactive visualisation of social research networks combined to semantic networks based on self-declared keywords provided by researchers. An other example closer to what we developed is Gargantext\footnote{\url{https://gargantext.org/}} that provides corpus exploration functionalities. Linkage\footnote{\url{https://linkage.fr/}} is a similar tool with different methods, using latent topic allocation for networks with textual annotations~\citep{bouveyron2016stochastic}. We differentiate from these by exploring simultaneously multiple dimensions of semantic classification and more importantly by adding the geographical aspect. Furthermore, in comparison to various tools that private publishers are beginning to introduce, the open and collaborative nature of our work is crucial. For example, \cite{bohannon2014google} suggests that one must stay careful when using search results from a popular academic search engine, as the mechanisms of the ranking algorithm and thus the multiple biases are unknown. The comparison is similar with text-mining paying services provided by private companies, as we suggest that a subtle synergy between knowledge content and knowledge production processes (to which closed tools are an obstacle) can be more beneficial to both.

We have studied a scientific corpus of a journal in Geography, combining multiple points of view through their embedding in the geographical space. This work is therefore in itself reflexive, illustrating the kind of new approach to science it aims at promoting. We believe that the open tools we develop in this context will contribute to the empowerment of authors within Open Science.






\section*{Supplementary Material}

\subsection*{Statistics of authorship}

\begin{figure}
	\includegraphics[width=\linewidth]{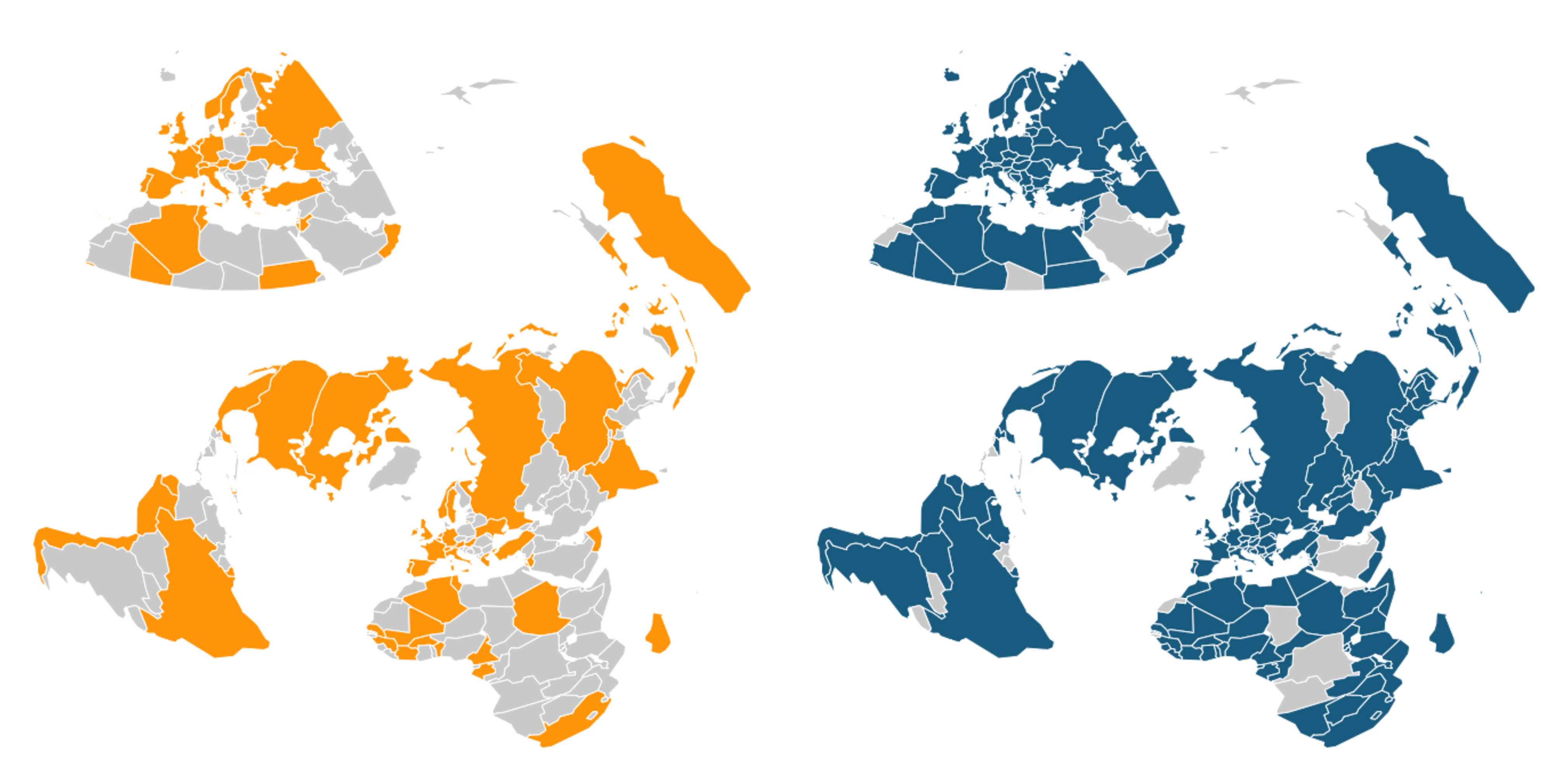}
\caption{\textit{(Left)} Countries with at least one author affiliated to an institution belonging to the country | 1996-2015. \textit{(Right)} Countries studied at least once | 1996-2015  } 
\label{fig:authoring}
\end{figure}

Most contributions come from French institutions (561), although French-speaking countries (35 papers from authors affiliated in Canada, 21 in Switzerland) and neighbouring countries (UK: 23 contributions, Italy: 18) are well represented too (fig. \ref{fig:authoring}, left). The geographical subjects of the articles themselves show a larger diversity, as the world is almost fully covered (fig. \ref{fig:authoring}, right). However, France and neighbouring countries such as Spain and Germany are the main focus of the majority of articles, although the United States are the 5th most studied single country.

Furthermore, the temporal evolution shows an accelerated growth of the number of authors, although the number of articles by 5-year period remains stable; a spread of geographical coverage, with more articles published about emerging countries and extra-European territories; along with a growing connexion in citation networks. There is a reinforcing bias towards a French-speaking authorship, revealed by the origin of authors as well as by the share of papers published in French.

\subsection*{Methodological details}

\subsubsection*{Internal semantic network}

The vertices (keywords) of the semantic network are described by two variables: frequency and degree. The frequency is the number of articles using the keyword. The degree is the total degree of the vertices in the network, that is, the number of edges linking a given keyword to the others (there is no distinction between in- and out- degree as the network is undirected). Both variables are distinct but correlated.
  
The edges are described by three variables: observed weight, expected weight and modal weight. For two given keywords the observed weight is the number of articles citing both keywords. The expected weight is the probability that the edge exists considering only the vertices' degree:

\begin{equation}
\nonumber
  P_{i \rightarrow j} = \frac{w_i w_j}{w(w - w_i)} ~~~~~~~  P_{j \rightarrow i} = \frac{w_i w_j}{w(w - w_j)}
\end{equation}

\begin{equation}
\nonumber
P_{i \leftrightarrow j} = P_{i \rightarrow j} \cup P_{j \rightarrow i}
\end{equation}

\begin{equation}
\nonumber
w_{i \leftrightarrow j}^e = \frac{w}{2} P_{i \leftrightarrow j}
\end{equation}

The probability of a link between $i$ and $j$ ($P_{i \rightarrow j}$) is defined as the cross-product of the marginal sums ($w_i$ and $w_j$) divided by the total weight ($w$). This can be seen as a quasi-modularity measure or a quasi chi-squared distance. The only difference is the null diagonal that creates asymmetric probabilities. The expected weight ($w_{i \leftrightarrow j}^e$) is the product of the probability and the mid-sum of weights.

Eventually the \textit{modal weight} is computed as a ratio between the observed weight and the squared-root expected weight of the edge (such as a Pearson residual in a chi-square analysis of a contingency table). This modal weight can be used as a preferential attachment measure. 

Based on this preferential attachment measure, two kinds of visualisations are proposed: semantic fields and communities. The semantic field shows for any given keyword at the centre of the plot all its neighbours at a distance inversely proportional to the modal weight.

\subsubsection*{External semantic network}

Having constructed the citation neighborhood, we introduce a method to analyse its content through text mining. More precisely, we focus on the \emph{relevant} keywords of abstracts, in a precise sense, which was introduced by~\cite{chavalarias2013phylomemetic} to study the evolution of scientific fields, and later refined and scaled to big data on a patent database by~\cite{bergeaud2017classifying}. Using co-occurrences of $n$-grams (keywords with multiple components, obtained after a first text cleaning and filtering), the deviation from an uniform distribution across texts using a chi-squared test gives a measure of keyword relevance, on which a fixed number $N_k = 50,000$ is filtered. The weighted co-occurrence network between relevant keywords captures their second-order relationship and we assume that its topology contains information on the structure of disciplines that are present in the citation network. We proceed to a sensitivity analysis of the network community structure to filtering parameters (minimal edge weight $\theta_w$, maximal node degree $k_{max}$). We choose parameters giving a Pareto optimal solution for modularity and network size (see \citep{raimbault2017exploration} for more details, as we follow exactly the same procedure).

\subsubsection*{Topics allocation}

Topic classification of text documents is an intense field of research, which has developed several algorithms. In this field, a topic is considered as a set of words frequently used together in the same document, and a text document as a mixture of topics. Following a long standing development in natural language processing from the weighting scheme of words called Term frequency-inverse document frequency (\textit{tfidf}) introduced by \cite{salton_introduction_1986} to first generative probabilistic model of \cite{hofmann1999probabilistic}, \cite{blei2003latent} have proposed a subsequent evolution with LDA.

The LDA method considers texts in a destructured way, i.e. words proximity or words presence in a same sentence are irrelevant. Articles thus become bags of words. To alleviate the disadvantage of destructuring the text, different methods can be used. The probabilistic tagging method proposed by \cite{schmid_probabilistic_1994} and used in this article aims at characterising each word by its function in the sentence, allowing us to filter only nouns, articles and verbs. The tagging includes also the transformation from plural forms into singular, and from conjugated verb form into infinitive. Each word is then associated with a frequency in a document, which can be weighted using the \textit{tfidf} weights. Out of many options, we use the form of \textit{tfidf} given by
\[
\textrm{\textit{tfidf}}_{t,d,D} = f _{t,d} \cdot log\left(\frac{N}{ |d \in D : t \in d| }\right)
\]

where $f_{t,d}$ is the frequency of the term $t$  in the document $d$, $N$ is the total number of documents in the corpus and $|d \in D : t \in d|$ is the number of documents $d$ in a reduced corpus $D$ where the term $t$ appears. This way, after having destructured text documents, filtered only nouns, articles and verbs, and finally weighted each word, we produce the matrix of weights per document and word in terms of topics, using the LDA model. 

\begin{figure} 
	\includegraphics[width=\linewidth]{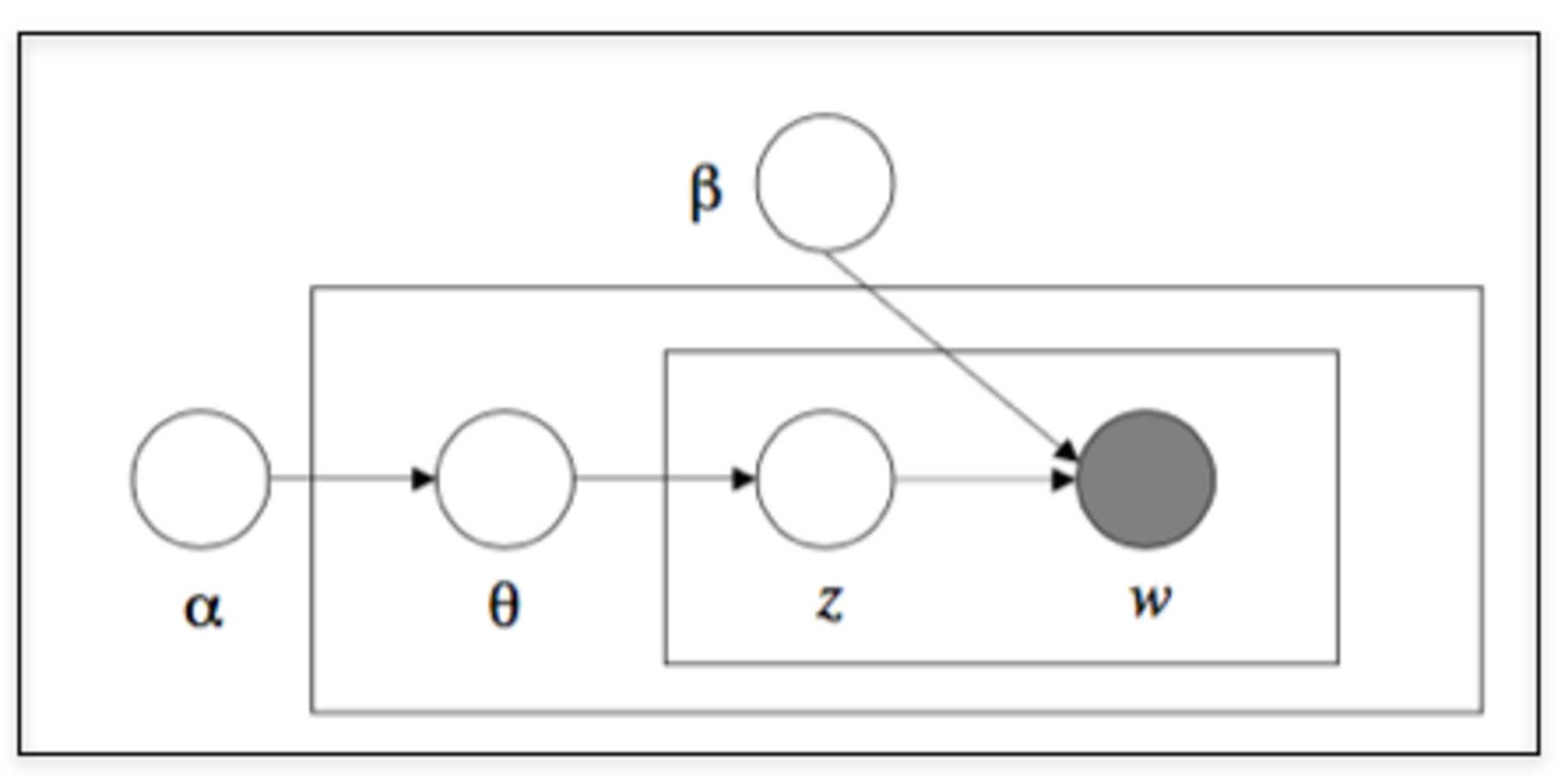}
\caption{Graphical model representation of the Latent Dirichlet allocation: a box symbolizes a level of the model (corpus, document, word); a blank disk, a parameter; the grey disk, the product of the generative model, i.e. a word drawn for a document of the whole corpus; an arrow, a dependency, i.e. \(\theta\) is drawn from a distribution with parameter \(\alpha\). See the section dedicated to the method for more explanations. Source: \citep{blei2003latent}.} 
\label{fig:lda} 
\end{figure}

LDA is a Bayesian hierarchical model (fig. \ref{fig:lda}). We give details of its structure in the following. This model considers three levels: corpus, document and word. Each level is defined by a set of probabilistic distributions and their parameters. Then, at the corpus level, the model has parameters \(\alpha\) and \(\beta\). \(\alpha\) is a vector of positive real numbers (one per topic). \(\beta\) is a matrix describing the probability of each word of the dictionary (columns) to be included in each topic (row). At the document level, the generative process begins by drawing the number of words from a Poisson distribution with parameter \(\epsilon\) and a vector \(\theta\) from a Dirichlet distribution with parameter \(\alpha\). Finally, at the word level, the topic \(z\) of the word is drawn from a multinomial distribution with parameter \(\theta\) and a word is drawn from a multinomial distribution with parameter the vector probability line of the topic \(z\) from the parameter matrix \(\beta\). All these parameters are estimated here using a Markov Chain Monte Carlo algorithm, called a Gibbs sampler by \cite{geman_stochastic_1984}. This algorithm generates a vast amount of draws of each parameters needed by the generative model, and produces a distribution of the value of each parameter. The main product is the \(\beta\) parameter, which is the probability of a word per topic, which is then analysed in order to understand the topic found in the corpus.

Throughout this process, the number of topics remains a fixed parameter. In order to select an optimal number, \cite{blei2003latent} proposed a graphical method aiming to identify the number of topics with a minimal perplexity and a maximal entropy. The perplexity is a measurement of how well a probability model predicts the terms used in each document of a corpus. A lower perplexity indicates a better generalization ability of the model.

\subsection*{Results}

\subsubsection*{Topic allocation}

\begin{figure}
	\includegraphics[width=\linewidth]{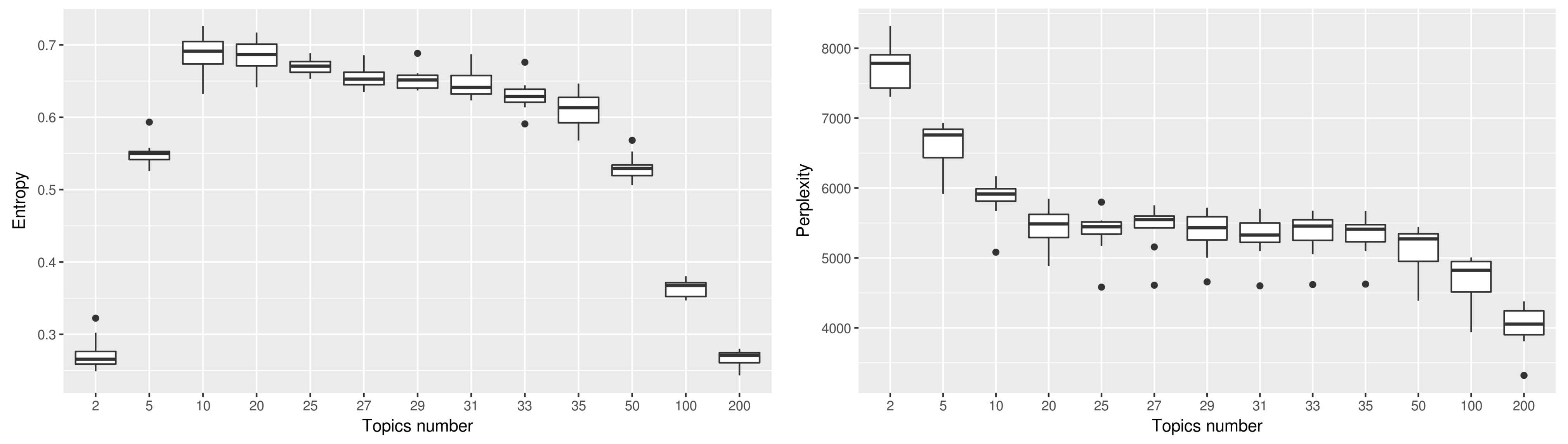}
\caption{\textit{(Left)} Entropy of the LDA model per number of topics. \textit{(Right)} Perplexity of the LDA model per number of topics.} 
\label{fig:entropy-perplexity}
\end{figure}

To choose the number of topics, we estimated the LDA model with the following number of topics: 2, 5, 10, 20, 25, 27, 29, 31, 33, 35, 50, 100 and 200. We test the robustness of results to stochasticity by iterating ten times each parameter estimation for a specific number of topics. In figure \ref{fig:entropy-perplexity} (left), the number of topics with a maximal entropy are 10 or 20. In figure \ref{fig:entropy-perplexity} (right), the lowest perplexity is reached with 20 topics, which becomes the optimal number of topics.

We give in table~\ref{tab:words-list} the content of the 20 topics for the classification chosen, in terms of keywords ordered by probability of appearance in each topic.

\begin{table}
\caption{List of top words for each topic. The french words are translated into english. The number indicates the probability of the word to belong to the topic, i.e. the word "district" in topic 1 has a probability of 8\% to be drawn for a document owning this topic. We present only words having a probability greater than 2\%.\medskip}
\label{tab:words-list}
\begin{tabular}{c|p{110mm}}
\hline
  Index & Words (probability) list \\
  \hline\footnotesize
  1 & district (8), city (7), housing (5), household (3) \\
  2 & move (6), mobility (5), density (4), accessibility (3), indicator (3), simulation (3), modeling (2), scenario (2) \\
  3 & image (8), soil (7), occupation (4), surface (4), vegetation (2), map (2), resolution (2), pixel (2), landscape (2)\\
  4 & planning (7), governance (4), urban planning (2), sustainability (2), document (2), participation (2) \\
  5 & risk (9), vulnerability (5), hazard (3), flood (2), city (2), water (2), disaster (2), management (2) \\
  6 & firm (7), healthcare (3), care (2) \\
  7 & the (16), and (8), The (2), for (2), are (2) \\
  8 & index (4), agent (3), graph (2), vertex (2), mountain (2) \\
  9 & city (5) \\
  10 & water (8), exploitation (4), management (3), farmer (2), agriculture (2), parcel (2) \\
  11 & map (10), cartography (3), journal (3), http (2), image (2), atlas (2) \\
  12 & geography (8), geographer (3), author (3), document (3), science (2) \\
  13 & island (5), student (3), geography (3), education (2), identity (2), image (2), university (2) \\
  14 & city (23), agglomeration (4), metropolis (3), area (2), urbanization (2) \\
  15 & Border (3), China (2), State (2), States (2), Brazil (2), Asia (2) \\
  16 & vote (5), map (3), party (2) \\
  17 & village (10), Pole (3), Departement (3), Area (2), Map (2), University (2), Student (2) \\
  18 & village (5), season (2), rain (2), resort (2), valley (2), precipitation (2), speed (2) \\
  19 & landscape (9), heritage (2), image (2), tourist (2) \\
  20 & port (4), sea (3), wind (2), station (2), breeze (2), Tunis (2), temperature (2) \\\hline
\end{tabular}
\end{table}

\subsection*{Complementarity of methods}

\subsubsection*{Flow diagrams}

We describe here the method used to quantify the overlap between classifications. If a method $M_1$ (for ex. based on citation communities) is composed of $n$ categories and a method $M_2$ (for ex. based on keywords communities) is composed of $m$ categories, we compute for each article $n*m$ products of co-occurrences and then sum these products into $flows$ for the whole Cybergeo corpus.\\
If the methods were equivalent ways of describing and clustering articles, we would expect all the flows between communities to be $1:1$, $n:1$ or $1:n$, given that the methods do not give the same number of clusters. If the methods were completely orthogonal, we should find that each flow is proportional to the size of the origin cluster and of the size of the destination clusters. The fact that we find $n:n$ flows and that they are not determined entirely by the size of the clusters at origin and destination means that our three methods of semantic clustering are neither equivalent nor orthogonal (Figure \ref{fig:complementarity}). On the contrary, they shed different lights on the journal corpus.\\

\begin{figure*}
	\includegraphics[width=\linewidth]{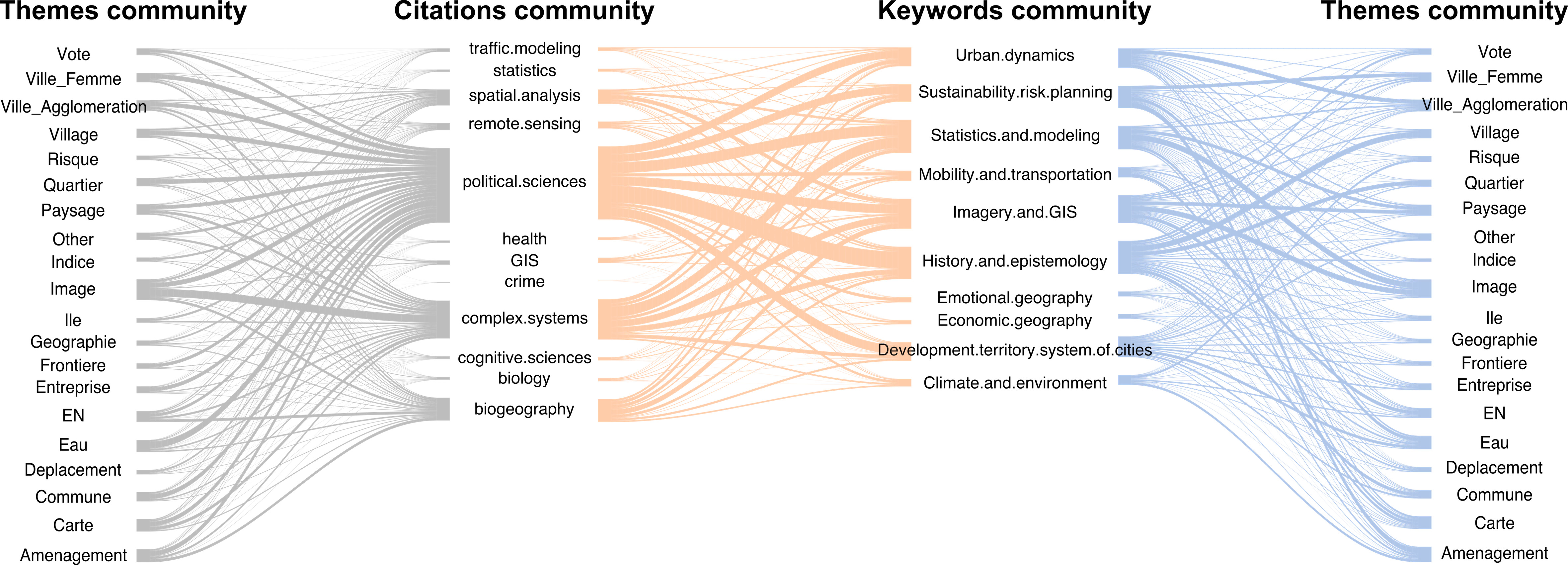}
\caption{Overlap between semantic communities} 
\label{fig:complementarity} 
\end{figure*}

\subsubsection*{Correlations between classifications}

We synthesize the flow relations between classifications by looking at their covariance structure in an aggregated way. More precisely, given the probability matrices $(p_{ki}) = (P_i)$ and $(p_{kj}) = (P_j)$ summarizing two classifications, where articles are indexed by rows, we estimate the correlation matrix between their columns $\rho_{ij} = \hat{\rho}\left[P_i,P_j\right]$ using a standard Pearson correlation estimator. We look then at aggregated measures, namely minimal correlation, maximal correlation and mean absolute correlation. In order to have a reference to interpret the values of these correlations, we compare them to two null models obtained by bootstrapping random corpuses. The estimate for the lower null model ($\rho_0$) is expected to minimize correlation and is obtained by shuffling all rows of one of the two matrices, which is done successively on both to ensure symmetry. The upper null model ($\rho_+$) is constructed by computing correlations between one matrix and the same where a fixed proportion of rows have been shuffled. We set this proportion to 50\%, which is a rather high level of similarity, and compute the model for both matrices each time. Average and standard deviations are computed for null models on $b=10000$ bootstrap repetitions. Table~\ref{tab:cors} summarizes the results. We find that the maximal correlation for the Cybergeo corpus, which can be interpreted as a maximum overlap between approaches of semantic clustering, is always significantly smaller (around $5\cdot \sigma$) than for the upper null model. This confirms that our three classifications are highly independent of one another in their main components. It is interesting to note that for Keywords/Themes, the mean absolute correlation is within the standard error range of the mean absolute correlation of the upper null model, suggesting that these two must be rather close on small overlaps. They are actually closer than with Citations for all indicators. We also confirm that Themes/Citations has the lowest mean absolute overlap.

\begin{table}
\begin{threeparttable}
\caption{\footnotesize{\textsc{Correlations between classifications}}}
\label{tab:cors}
\begin{tabular}{p{1.3cm}|ccccccccc}
 \hline\cr
 & $\min \rho$ & $\min \rho_0$ & $\min \rho_+$ & $\max \rho$ & $\max \rho_0$ &$\max \rho_+$ & $< \left| \rho \right|>$ & $< \left| \rho_0 \right|>$ & $< \left| \rho_+ \right|>$ \\\cr\hline\cr
 Themes & -0.30 &$-0.12$&$-0.17$&0.36&$0.21$&$0.69$&0.059&$0.043$&$0.073$\\
 /Citations &  &$\pm 0.019$&$\pm 0.071$&&$\pm 0.042$&$\pm 0.070$&&$\pm 0.0021$&$\pm 0.012$\\\cr\hline\cr
 Citations & -0.26 & $-0.096$ & $-0.20$ & 0.30 & $ 0.13$ & $0.64$ & 0.070 & $ 0.034$ & $ 0.092$\\
 /Keywords &  & $\pm 0.015$ & $\pm 0.047$ && $\pm 0.027 $ & $\pm0.068$ & & $\pm 0.0026 $ & $\pm 0.0081$\\\cr\hline\cr
 Keywords &-0.20&$-0.11$&$-0.13$&0.51&$0.17$&$0.66$&0.091&$0.040$&$0.080$\\
 /Themes & &$\pm0.013$&$\pm0.030$&&$\pm0.032$&$\pm0.075$&&$\pm0.0022$&$\pm0.020$\\\cr
 \hline
\end{tabular}
 \begin{tablenotes}
       \item \protect\scriptsize{\textbf{Notes}: For each pair of classification and measure, we also give average and standard deviation for lower ($\rho_0$) and upper ($\rho_+$) null models, obtained by bootstrapping $b=10000$ random corpuses.}
    \end{tablenotes}
  \end{threeparttable}
\end{table}

\begin{figure}
	\includegraphics[width=\linewidth]{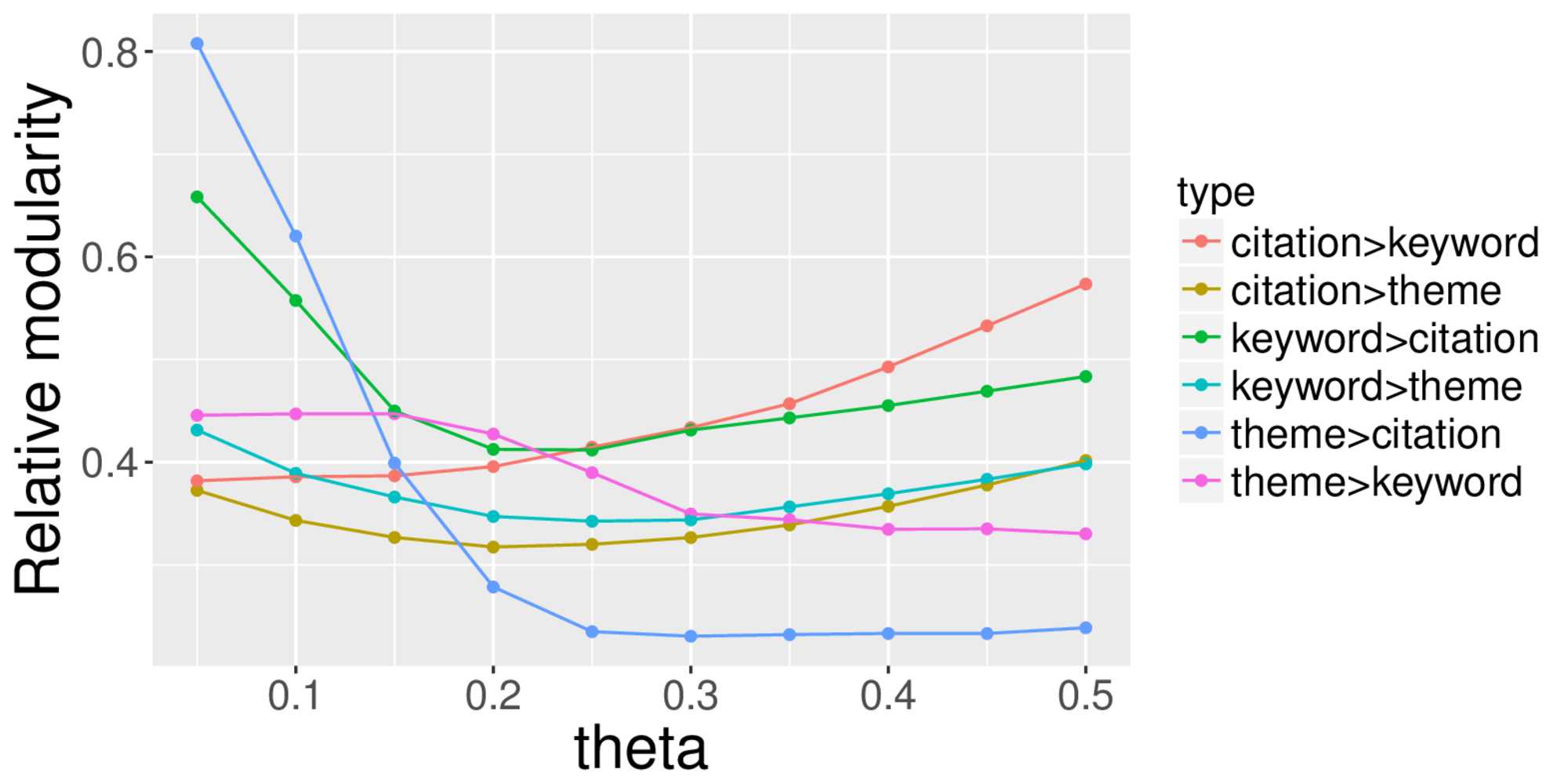}
\caption{Evaluation of the complementarity of classifications through network modularities. The plot gives the relative modularity of the first classification in the network induced by the second with the threshold $\theta$ (see text), for each couple of classifications (color).}
\label{fig:modularities}
\end{figure}

To make these conclusions more robust, we complement the analysis with a network modularity analysis, which is a widely applied method to evaluate the relevance of a classification within a network. To be able to compare two classifications, since the citation network is too sparse for any analysis as mentioned, we evaluate the modularity of a classification within the network induced by the other. More precisely, given a distance threshold $\theta$ and two documents given by their probabilities within a classification $\vec{p}_i^{(c)},\vec{p}_j^{(c)}$, we consider the network with documents as nodes linked if and only if $d(\vec{p}_i^{(c)},\vec{p}_j^{(c)})<\theta$ with $d$ euclidian distance. We can then compute the multi-class modularity of the other classification in the sense of~\cite{nicosia2009extending}. We show in figure~\ref{fig:modularities}, for different thresholds, the modularities normalized by the modularity of the network classification within its own network. The closest the measure is to 1, the closer are the classifications. Most of couple have low values for large ranges of $\theta$, confirming the previous conclusions of orthogonality. Furthermore, the different behavior as a function of $\theta$ (increasing or decreasing) suggests different \emph{internal structures} of classification, what is consistent with the fact that they rely on different processes to classify data.

\begin{thebibliography}{39}
\providecommand{\natexlab}[1]{#1}
\providecommand{\url}[1]{\texttt{#1}}
\providecommand{\urlprefix}{URL }
\expandafter\ifx\csname urlstyle\endcsname\relax
  \providecommand{\doi}[1]{DOI:\discretionary{}{}{}#1}\else
  \providecommand{\doi}{DOI:\discretionary{}{}{}\begingroup
  \urlstyle{rm}\Url}\fi

\bibitem[{Ackermann et~al.(2003)Ackermann, Mering and
  Quensiere}]{ackermann2003analysis}
Ackermann G, Mering C and Quensiere J (2003) Analysis of built-up areas
  extension on the {P}etite {C}{\^o}te region ({S}enegal) by remote sensing.
\newblock \emph{Cybergeo: European Journal of Geography} 9(249).

\bibitem[{B{\'e}lizal et~al.(2011)B{\'e}lizal, Lavigne and
  Grancher}]{belizal2011quand}
B{\'e}lizal {\'E}d, Lavigne F and Grancher D (2011) Quand l'al{\'e}a devient la
  ressource: l'activit{\'e} d'extraction des mat{\'e}riaux volcaniques autour
  du volcan {M}erapi ({I}ndon{\'e}sie) dans la compr{\'e}hension des risques
  locaux.
\newblock \emph{Cybergeo: European Journal of Geography} 23555.

\bibitem[{Blondel et~al.(2008)Blondel, Guillaume, Lambiotte and
  Emmanuel}]{blondel_fast_2008}
Blondel VD, Guillaume JL, Lambiotte R and Emmanuel L (2008) Fast unfolding of
  communities in large networks.
\newblock \emph{Journal of Statistical Mechanics: Theory and Experiment} (10):
  10008.

\bibitem[{Bohannon(2014)}]{bohannon2014google}
Bohannon J (2014) Google scholar wins raves - but can it be trusted?
\newblock \emph{American Association for the Advancement of Science} .

\bibitem[{B{\"o}rner et~al.(2015)B{\"o}rner, Theriault and
  Boyack}]{borner2015mapping}
B{\"o}rner K, Theriault TN and Boyack KW (2015) Mapping science introduction:
  past, present and future.
\newblock \emph{Bulletin of the Association for Information Science and
  Technology} 41(2): 12--16.

\bibitem[{Bouveyron et~al.(2016)Bouveyron, Latouche and
  Zreik}]{bouveyron2016stochastic}
Bouveyron C, Latouche P and Zreik R (2016) The stochastic topic block model for
  the clustering of vertices in networks with textual edges.
\newblock \emph{Statistics and Computing} : 1--21.

\bibitem[{Citron and Way(2018)}]{CITRON2018181}
Citron DT and Way SF (2018) {Network assembly of scientific communities of
  varying size and specificity}.
\newblock \emph{Journal of Informetrics} .

\bibitem[{Cronin and Sugimoto(2014)}]{cronin2014beyond}
Cronin B and Sugimoto CR (2014) \emph{Beyond bibliometrics: Harnessing
  multidimensional indicators of scholarly impact}.
\newblock MIT Press.

\bibitem[{Devaux et~al.(2007)Devaux, Fotsing and
  Ch{\'e}ry}]{devaux2007extraction}
Devaux N, Fotsing JM and Ch{\'e}ry JP (2007) Extraction automatique
  d'habitations en milieu rural de {PED} {\`a} partir de donn{\'e}es {THRS}.
\newblock \emph{Cybergeo: European Journal of Geography} 12581.

\bibitem[{Escobar et~al.(2000)Escobar, Polley and
  Williamson}]{escobar2000distribution}
Escobar FJ, Polley IS and Williamson IP (2000) Distribution of online
  cartographic products in {A}ustralia.
\newblock \emph{Cybergeo: European Journal of Geography} 168.

\bibitem[{Fecher and Friesike(2014)}]{fecher2014open}
Fecher B and Friesike S (2014) Open science: one term, five schools of thought.
\newblock In: \emph{Opening science}. Springer, pp. 17--47.

\bibitem[{Fortunato et~al.(2018)Fortunato, Bergstrom, B{\"o}rner, Evans,
  Helbing, Milojevi{\'c}, Petersen, Radicchi, Sinatra, Uzzi
  et~al.}]{fortunato2018science}
Fortunato S, Bergstrom CT, B{\"o}rner K, Evans JA, Helbing D, Milojevi{\'c} S,
  Petersen AM, Radicchi F, Sinatra R, Uzzi B et~al. (2018) Science of science.
\newblock \emph{Science} 359(6379): eaao0185.

\bibitem[{Frenken et~al.(2009)Frenken, Hardeman and
  Hoekman}]{frenken2009spatial}
Frenken K, Hardeman S and Hoekman J (2009) Spatial scientometrics: Towards a
  cumulative research program.
\newblock \emph{Journal of informetrics} 3(3): 222--232.

\bibitem[{Hicks et~al.(2015)Hicks, Wouters, Waltman, Rijcke and
  Rafols}]{hicks2015bibliometrics}
Hicks D, Wouters P, Waltman L, Rijcke Sd and Rafols I (2015) Bibliometrics: the
  leiden manifesto for research metrics .

\bibitem[{Jacobs(1961)}]{jacobs1961death}
Jacobs J (1961) The death and life of great american cities--vintage books.
\newblock \emph{New York} .

\bibitem[{Jelokhani-Niaraki and Malczewski(2012)}]{jelokhani2012web}
Jelokhani-Niaraki M and Malczewski J (2012) A web 3.0-driven collaborative
  multicriteria spatial decision support system.
\newblock \emph{Cybergeo: European Journal of Geography} 25514.

\bibitem[{Karami et~al.(2018)Karami, Dahl, Turner-McGrievy, Kharrazi and
  Shaw}]{KARAMI20181}
Karami A, Dahl AA, Turner-McGrievy G, Kharrazi H and Shaw G (2018)
  Characterizing diabetes, diet, exercise, and obesity comments on twitter.
\newblock \emph{International Journal of Information Management} 38(1): 1 -- 6.
\newblock \doi{https://doi.org/10.1016/j.ijinfomgt.2017.08.002}.
\newblock
  \urlprefix\url{http://www.sciencedirect.com/science/article/pii/S0268401217306126}.

\bibitem[{Kosmopoulos(2002)}]{kosmopoulos2002cybergeo}
Kosmopoulos C (2002) Cybergeo et les p{\'e}riodiques {\'e}lectroniques
  scientifiques.
\newblock \emph{Cybergeo: European Journal of Geography} 14253.

\bibitem[{Le~N{\'e}chet(2011)}]{le2011consommation}
Le~N{\'e}chet F (2011) Consommation d'{\'e}nergie et mobilit{\'e} quotidienne
  selon la configuration des densit{\'e}s dans 34 villes europ{\'e}ennes.
\newblock \emph{Cybergeo: European Journal of Geography} 23634.

\bibitem[{Livingstone(1995)}]{livingston_spaces_1995}
Livingstone DN (1995) The spaces of knowledge: contributions toward a
  historical geography of science.
\newblock \emph{{{Environment and planning D}}} 13: 13--42.

\bibitem[{Livingstone(2003)}]{livingston_science_2003}
Livingstone DN (2003) \emph{Puttings science in its place: geographies of
  scientific knowledge}.
\newblock Chicago: The University of Chicago Press.

\bibitem[{Lusso(2009)}]{lusso2009musees}
Lusso B (2009) Les mus{\'e}es, un outil efficace de r{\'e}g{\'e}n{\'e}ration
  urbaine? les exemples de {M}ons ({B}elgique), {E}ssen ({A}llemagne) et
  {M}anchester ({R}oyaume-{U}ni).
\newblock \emph{Cybergeo: European Journal of Geography} 21253.

\bibitem[{Newman(2014)}]{2013arXiv1310.8220N}
Newman M (2014) Prediction of highly cited papers.
\newblock \emph{EPL (Europhysics Letters)} 105(2): 28002.

\bibitem[{Niebles et~al.(2008)Niebles, Wang and
  Fei-Fei}]{niebles_unsupervised_2008}
Niebles JC, Wang H and Fei-Fei L (2008) Unsupervised {Learning} of {Human}
  {Action} {Categories} {Using} {Spatial}-{Temporal} {Words}.
\newblock \emph{International Journal of Computer Vision} 79(3): 299--318.
\newblock \doi{10.1007/s11263-007-0122-4}.
\newblock
  \urlprefix\url{https://link.springer.com.proxy.bnl.lu/article/10.1007/s11263-007-0122-4}.

\bibitem[{Omodei et~al.(2017)Omodei, De~Domenico and
  Arenas}]{2016arXiv160106075O}
Omodei E, De~Domenico M and Arenas A (2017) Evaluating the impact of
  interdisciplinary research: A multilayer network approach.
\newblock \emph{Network Science} 5(2): 235--246.

\bibitem[{Ozer and De~Longueville(2005)}]{ozer2005tsunami}
Ozer P and De~Longueville F (2005) Tsunami en {A}sie du {S}ud-{E}st: retour sur
  la gestion d'un cataclysme naturel apocalyptique.
\newblock \emph{Cybergeo: European Journal of Geography} 3081.

\bibitem[{Pumain(1996)}]{pumain1996cyberjournals}
Pumain D (1996) Are cyberjournals a promised land for researchers?
\newblock \emph{Environment and Planning A} (6): 951--954.

\bibitem[{Pumain(2001)}]{pumain2001cybergeo}
Pumain D (2001) Cybergeo, premi{\`e}re revue {\'e}lectronique fran{\c{c}}aise
  en g{\'e}ographie.
\newblock \emph{Comprendre les usages de l'Internet, Paris, {E}ditions {ENS}
  rue d'{U}lm} : 89--97.

\bibitem[{Putra and Baier(2009)}]{putra2009einfluss}
Putra DP and Baier K (2009) Der {E}influss ungesteuerter {U}rbanisierung auf
  die {G}rundwasserressourcen am {B}eispiel der indonesischen {M}illionenstadt
  {Y}ogyakarta.
\newblock \emph{Cybergeo: European Journal of Geography} 22573.

\bibitem[{Querriau et~al.(2004)Querriau, Kissiyar, Peeters and
  Thomas}]{querriau2004localisation}
Querriau X, Kissiyar M, Peeters D and Thomas I (2004) Localisation optimale
  d'unit{\'e}s de soins dans un pays en voie de d{\'e}veloppement: analyse de
  sensibilit{\'e}.
\newblock \emph{Cybergeo: European Journal of Geography} 3316.

\bibitem[{Raimbault(2019)}]{raimbault2017exploration}
Raimbault J (2019) Exploration of an interdisciplinary scientific landscape.
\newblock \emph{Scientometrics} : 1--25.

\bibitem[{Ross-Hellauer(2017)}]{10.12688/f1000research.11369.1}
Ross-Hellauer T (2017) What is open peer review? a systematic review [version
  1; referees: 1 approved, 2 approved with reservations].
\newblock \emph{F1000Research} 6(588).
\newblock \doi{10.12688/f1000research.11369.1}.

\bibitem[{Roth and Cointet(2010)}]{roth_social_2010}
Roth C and Cointet JP (2010) Social and semantic coevolution in knowledge
  networks.
\newblock \emph{Social Networks} 32(1): 16--29.

\bibitem[{Santamaria(2009)}]{santamaria2009schema}
Santamaria F (2009) Le sch{\'e}ma de d{\'e}veloppement de l'espace
  communautaire (sdec): application d{\'e}faillante ou {\'e}laboration
  probl{\'e}matique?
\newblock \emph{Cybergeo: European journal of geography} 22354.

\bibitem[{Shibata et~al.(2008)Shibata, Kajikawa, Takeda and
  Matsushima}]{shibata2008detecting}
Shibata N, Kajikawa Y, Takeda Y and Matsushima K (2008) Detecting emerging
  research fronts based on topological measures in citation networks of
  scientific publications.
\newblock \emph{Technovation} 28(11): 758--775.

\bibitem[{Vall{\'e}e(2009)}]{vallee2009disparites}
Vall{\'e}e J (2009) Les disparit{\'e}s spatiales de sant{\'e} en ville:
  l'exemple de {V}ientiane ({L}aos).
\newblock \emph{Cybergeo: European Journal of Geography} 22775.

\bibitem[{Wen et~al.(2017)Wen, Horlings, van~der Zouwen and Van~den
  Besselaar}]{wen2017mapping}
Wen B, Horlings E, van~der Zouwen M and Van~den Besselaar P (2017) Mapping
  science through bibliometric triangulation: An experimental approach applied
  to water research.
\newblock \emph{Journal of the Association for Information Science and
  Technology} 68(3): 724--738.

\bibitem[{Wilson et~al.(2017)Wilson, Bryan, Cranston, Kitzes, Nederbragt and
  Teal}]{wilson2017good}
Wilson G, Bryan J, Cranston K, Kitzes J, Nederbragt L and Teal T (2017) Good
  enough practices in scientific computing.
\newblock \emph{PLoS Comput Biol} 13(6): e1005510.

\bibitem[{Withers(2009)}]{withers_place_2009}
Withers CWJ (2009) Place and the spatial turn in geography and history.
\newblock \emph{Journal of the History of Ideas} 70: 637--658.


\bibitem[{Bergeaud et~al.(2017)Bergeaud, Potiron and
  Raimbault}]{bergeaud2017classifying}
Bergeaud A, Potiron Y and Raimbault J (2017) Classifying patents based on their
  semantic content.
\newblock \emph{PloS one} 12(4): e0176310.

\bibitem[{Blei et~al.(2003)Blei, Ng and Jordan}]{blei2003latent}
Blei DM, Ng AY and Jordan MI (2003) Latent dirichlet allocation.
\newblock \emph{Journal of machine Learning research} 3(Jan): 993--1022.

\bibitem[{Chavalarias and Cointet(2013)}]{chavalarias2013phylomemetic}
Chavalarias D and Cointet JP (2013) Phylomemetic patterns in science evolution
  - the rise and fall of scientific fields.
\newblock \emph{PloS one} 8(2): e54847.

\bibitem[{Geman and Geman(1984)}]{geman_stochastic_1984}
Geman S and Geman D (1984) Stochastic {Relaxation}, {Gibbs} {Distributions},
  and the {Bayesian} {Restoration} of {Images}.
\newblock \emph{IEEE Trans. Pattern Anal. Mach. Intell.} 6(6): 721--741.
\newblock \doi{10.1109/TPAMI.1984.4767596}.
\newblock \urlprefix\url{http://dx.doi.org/10.1109/TPAMI.1984.4767596}.

\bibitem[{Hofmann(1999)}]{hofmann1999probabilistic}
Hofmann T (1999) Probabilistic latent semantic indexing.
\newblock In: \emph{Proceedings of the 22nd annual international ACM SIGIR
  conference on Research and development in information retrieval}. ACM, pp.
  50--57.

\bibitem[{Nicosia et~al.(2009)Nicosia, Mangioni, Carchiolo and
  Malgeri}]{nicosia2009extending}
Nicosia V, Mangioni G, Carchiolo V and Malgeri M (2009) Extending the
  definition of modularity to directed graphs with overlapping communities.
\newblock \emph{Journal of Statistical Mechanics: Theory and Experiment}
  2009(03): P03024.

\bibitem[{Raimbault(2019)}]{raimbault2017exploration}
Raimbault J (2019) Exploration of an interdisciplinary scientific landscape.
\newblock \emph{Scientometrics} : 1--25.

\bibitem[{Salton and McGill(1986)}]{salton_introduction_1986}
Salton G and McGill MJ (1986) \emph{Introduction to {Modern} {Information}
  {Retrieval}}.
\newblock New York, NY, USA: McGraw-Hill, Inc.
\newblock ISBN 978-0-07-054484-0.

\bibitem[{Schmid(1994)}]{schmid_probabilistic_1994}
Schmid H (1994) Probabilistic {Part}-of-{Speech} {Tagging} {Using} {Decision}
  {Trees}.
\newblock Manchester, United Kingdom.



\end{thebibliography}
\end{document}